\documentclass[%
 reprint,
superscriptaddress,
 amsmath,amssymb,
 aps,
prb,
]{revtex4-2}

\usepackage{graphicx}
\usepackage{dcolumn}
\usepackage{bm}
\usepackage{multirow}
\usepackage{color}

\usepackage[normalem]{ulem} 

\def\Vec#1{{\bf #1}}
\def\GVec#1{\mbox{\boldmath $#1$}}

\begin{document}

\preprint{APS/123-QED}

\title{Quasicrystalline electronic states in twisted bilayers\\ and the effects of interlayer and sublattice symmetries}

\author{J.\,A.\,Crosse}
\affiliation{Arts and Sciences, NYU Shanghai, Shanghai, China}
\affiliation{NYU-ECNU Institute of Physics at NYU Shanghai, Shanghai, China}

\author{Pilkyung Moon}%
 \email{Corresponding author: pilkyung.moon@nyu.edu}
\affiliation{Arts and Sciences, NYU Shanghai, Shanghai, China}
\affiliation{NYU-ECNU Institute of Physics at NYU Shanghai, Shanghai, China}
\affiliation{Department of Physics, New York University, New York, USA}
\affiliation{State Key Laboratory of Precision Spectroscopy, East China Normal University, Shanghai, China}

\date{\today}

\begin{abstract}
We study the electronic structure of
quasicrystals composed of incommensurate stacks of
atomic layers. We consider two systems: a pair of square lattices with a relative twist angle of $\theta=45^\circ$ and a pair of hexagonal lattices with a relative twist angle of $\theta=30^\circ$, with various interlayer interaction strengths.
This constitutes every two-dimensional bilayer quasicrystal system.
We investigate the resonant coupling governing the
quasicrystalline order in each quasicrystal symmetry,
and calculate the quasi-band dispersion.
The resonant interaction emerges in bilayer quasicrystals
if all the dominant interlayer interactions occur
between the atomic orbitals that have the same magnetic quantum number.
Thus, not only the quasicrystal composed of the widely studied graphene,
but also those composed of transition metal dichalcogenides
will exhibit the quasicrystalline states.
We find that some quasicrystalline states, which are usually obscured by
decoupled monolayer states,
are more prominent, i.e., "exposed",
in the systems with strong interlayer interaction.
We also show that
we can switch the states
between quasicrystalline configuration
and its layer components,
by turning on and off the interlayer symmetry.



\end{abstract}

\maketitle


\section{Introduction}

When two hexagonal lattices are overlapped, one
on top of the other at a twist angle $\theta=30^\circ$,
the atomic arrangement is mapped on to
a quasicrystalline lattice,
which is ordered but not periodic,
with a 12-fold rotational symmetry \cite{stampfli}.
Recently, it has been demonstrated that bilayer graphene with a precise rotation angle
of $30^\circ$ exhibits the atomic structures
satisfying the quasicrystalline tiling
as well as a spectrum respecting the 12-fold rotational symmetry
\cite{Ahn2018,suzuki2019ultrafast}.
Similar structures have also been realized
by growing bilayer graphene
on top of the Ni \cite{takesaki2016highly,yao2018quasicrystalline}
or Cu surface \cite{chen2016high,pezzini202030},
and also by a transfer method \cite{chen2016high}.

The conventional moir\'{e} effective theory,
which is based on the period of the moir\'{e} pattern
arising from the interference between the lattice periods,
cannot describe
the electronic structures of such
quasicrystals composed of incommensurate stack of atomic layers
(hereafter "vdW-QCs")
since the rotational symmetry of quasicrystals
does not commute with translation.
In our previous work, we developed a momentum-space tight-binding model
which can describe the electronic structures of
atomic layers stacked at any configuration
without relying on the moir\'{e} periodicity \cite{moon2019quasicrystalline}.
This model enabled us to 
reveal the quasi-band dispersion
and the emergence of the electronic states
having the quasicrystalline order in the vdW-QC
composed of two graphene layers stacked at $30^\circ$
by fully respecting the rotational symmetry of quasicrystals
as well as the translational symmetry of constituent layers.
While conventional quasicrystals can be viewed as \textit{intrinsic quasicrystals}
where all the atomic sites are intrinsically arranged
in the quasiperiodic order,
vdW-QCs are regarded as \textit{extrinsic quasicrystals},
in that they are composed of a pair of perfect crystals
having independent periodicities,
and the quasiperiodic nature appears
only in the perturbational coupling between the two subsystems.
Thus, vdW-QCs provide a unique opportunity to design
quasicrystalline states by using atomic layers
with various symmetries
and also to control the quasicrystalline interaction
by controlling the interlayer interaction.

In this paper, we numerically investigate
the electronic structures of vdW-QCs
for every possible rotational symmetry in two-dimensional space.
Since a periodic two-dimensional atomic layer
can have 2-, 4-, 6-fold rotational symmetry,
we can make only
8-fold [octagonal, Fig.~\ref{fig_atomic_structure_and_BZ}(a)]
or 12-fold [dodecagonal, Fig.~\ref{fig_atomic_structure_and_BZ}(d)]
vdW-QCs with two two-dimensional layers. This can be achieved 
by stacking two square lattices at $45^\circ$
or by stacking two hexagonal lattices at $30^\circ$, respectively.
We first find the resonant condition,
which gives quasicrystalline order to the electronic states,
in each system,
and show that such resonant interaction
emerges in this configuration
if all the dominant interlayer interactions
occur between the atomic orbitals that have the same
magnetic quantum number.
We calculate the quasi-band dispersion
for various interlayer interaction strength,
and identify the features which arise
from the quasicrystalline order as opposed to
those arising from the interaction common to any other $\theta$
in the spectrum of vdW-QCs.
In addition, we show that
some quasicrystalline states, which are usually obscured by additional weakly coupled states, are more prominent 
in vdW-QCs with strong interlayer interaction.
We also analyze the effects of lifting both interlayer and sublattice symmetry on the electronic structure,
and discuss the mixing between the quasicrystalline states,
which may influence the physical properties such as the optical selection rules.

The paper is organized as follows.
In Sec.~\ref{sec:theoretical_methods},
we present the atomic structures and
tight-binding model for vdW-QCs, and introduce the dual 
tight-binding approach in the momentum space.
And we reveal the resonant interaction
which gives the quasicrystalline electronic states
in vdW-QCs,
and investigate the effects of atomic orbitals that have different symmetries.
In Sec.~\ref{sec:octa} and \ref{sec:dodeca},
we derive the minimal Hamiltonian
and calculate the band dispersion and wave functions
of octagonal and dodecagonal vdW-QCs, respectively.
We also investigate the effects of
various interlayer interaction strengths,
the features arising from 2-wave mixing,
and the effects of the lifting of
interlayer and sublattice potential asymmetry.
A brief conclusion is given in
Sec.~\ref{sec:conclusions}.

\section{\label{sec:theoretical_methods}Theoretical methods}

\subsection{Atomic structure and Brillouin zones of quasicrystalline twisted bilayers}

\begin{figure*}
	\begin{center}
		\leavevmode\includegraphics[width=0.9\hsize]{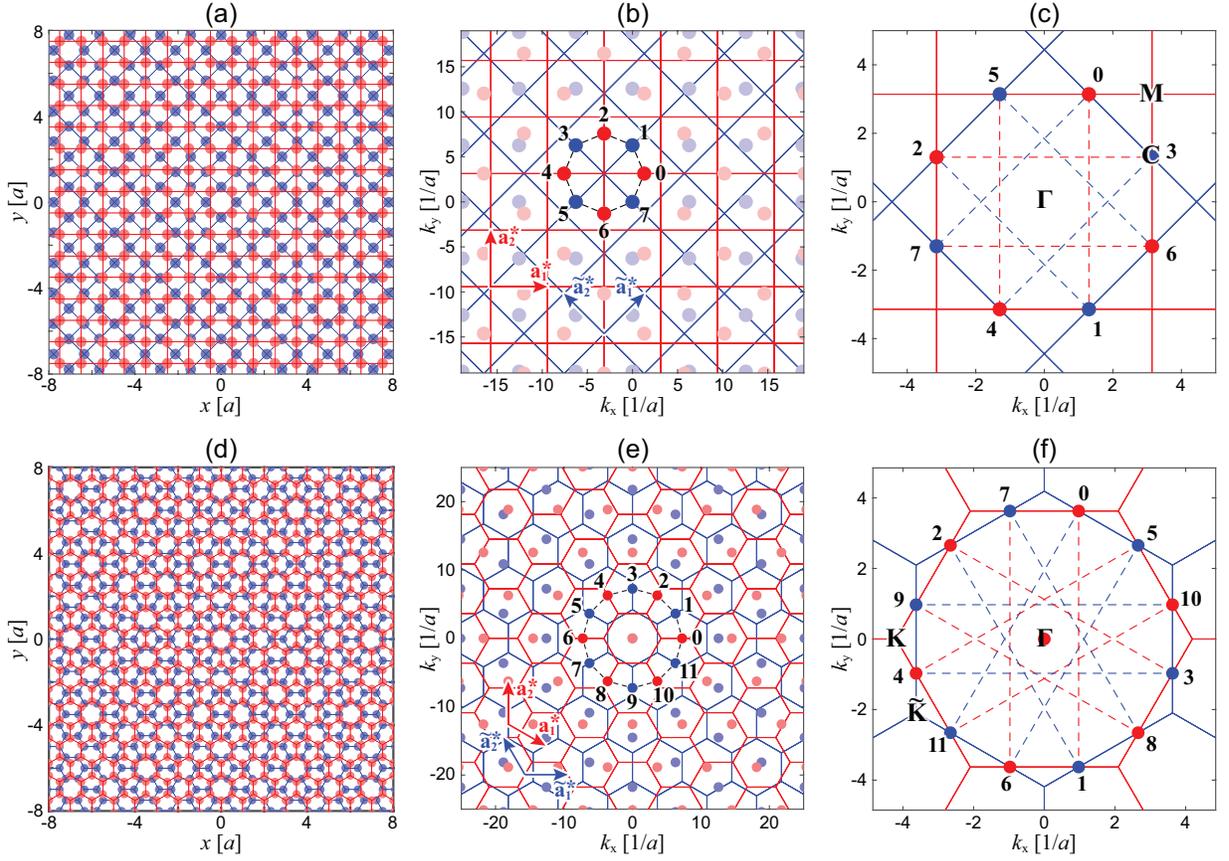}
	\end{center}
	\caption{
		(a) Real-space lattice structures of octagonal vdW-QCs.
		The red and blue squares represent the unit cells and the red and blue circles the atomic sites of layer 1 and 2, respectively.
		(b) Dual tight-binding lattice in the momentum space
		for octagonal vdW-QCs (see text).
		Red and blue squares show the extended Brillouin zones of layer 1 and 2, respectively.
		The number $n$ represents the position of $\Vec{Q}_n$ $(n=0,1,2,\cdots, 7)$,
		and the dashed lines indicate the connections in the 8-ring effective Hamiltonian.
		The red circles represent the wave numbers $\Vec{k}$ for layer 1,
		and blue ones represent the inverted wave numbers $\hat{\Vec{k}} - \tilde{\Vec{k}}$ for layer 2,
		where $\hat{\Vec{k}}$ is taken as $\Vec{Q}_0$ here.
		(c) The wave numbers $\Vec{C}_n$ $(n=0,1,2,\cdots, 7)$ at the cross points between the first Brillouin zones
		of the two lattices,
		which are the original positions of $\Vec{k}$ (layer 1) and $\tilde{\Vec{k}}$ (layer 2) associated with $\Vec{Q}_n$.
		The dashed lines indicate the connections in the 8-ring Hamiltonian as in (b).
		Due to the symmetry, these 8 wave numbers are all degenerate in energy.
		(d), (e), (f) Plots similar to (a), (b), (c) for dodecagonal vdW-QCs.
		Here, $n$ for $\Vec{Q}_n$ and $\Vec{C}_n$ runs from 0 to 11,
		and $\hat{\Vec{k}}=\Vec{0}$.
		}
	\label{fig_atomic_structure_and_BZ}
\end{figure*}

We define the atomic structure of the octagonal vdW-QCs
by starting from two perfectly overlapping square lattices
and rotating the layer 2 around the center of the square by $\theta=45^\circ$
[Fig.~\ref{fig_atomic_structure_and_BZ}(a)].
We set $xy$ coordinates parallel to the layers and $z$ axis perpendicular to
them.
The system belongs to the symmetry group $D_{4d}$,
and it is invariant under an improper rotation $R(\pi/4) M_z$, 
where $R(\theta)$ is the rotation by an angle $\theta$ around $z$ axis,
and $M_z$ is the mirror reflection with respect to $xy$ plane. 
The primitive lattice vectors of layer 1 are taken as $\Vec{a}_1 = a(1,0)$ and $\Vec{a}_2 = a(0,1)$,
where $a$ is the lattice constant,
and those of the layer 2 as $\tilde{\Vec{a}}_i = R(\pi/4)\,\Vec{a}_i$.
In this paper, we model the square lattices
by a minimal, one orbital model with one sublattice site.
Then, the atomic positions are given by
\begin{align}
&\Vec{R}_{X}=n_1\Vec{a}_{1}+n_2\Vec{a}_{2}+\GVec{\tau}_X
&(\mbox{layer 1}), \nonumber\\
&\Vec{R}_{\tilde{X}}=\tilde{n}_1\tilde{\Vec{a}}_{1}+\tilde{n}_2\tilde{\Vec{a}}_{2}+\GVec{\tau}_{\tilde{X}}
&(\mbox{layer 2}),
\label{eq_atomic_position}
\end{align}
where $n_i$ and $\tilde{n}_i$ are integers, 
$X$ ($\tilde{X}$) denotes the sublattice site (only one in this case) of layer 1
(layer 2) of which position in the unit cell is defined
by $\GVec{\tau}_X=(a/2, a/2)$ [$\GVec{\tau}_{\tilde{X}}=R(\pi/4) \GVec{\tau}_X + d\Vec{e}_z$].
Here, $d$ is the interlayer spacing between the two layers and
$\Vec{e}_z$ is the unit vector normal to the layer.
The reciprocal lattice vectors
of layer 1 are given by $\Vec{a}^*_1 = (2\pi/a)(1,0)$ and $\Vec{a}^*_2=(2\pi/a)(0,1)$,
and layer 2 by $\tilde{\Vec{a}}^*_i = R(\pi/4)\, \Vec{a}^*_i$.
The red and blue squares in
Figs.~\ref{fig_atomic_structure_and_BZ}(b) and (c) show the Brillouin zones
of layer 1 and 2 in octagonal vdW-QCs, respectively.

Likewise, we define the atomic structure of the dodecagonal vdW-QCs
by starting from two perfectly overlapping hexagonal lattices
(i.e., AA-stacked bilayers)
and rotating the layer 2 around the center of the hexagon by $\theta=30^\circ$
[Fig.~\ref{fig_atomic_structure_and_BZ}(d)].
The system belongs to the symmetry group $D_{6d}$,
and it is invariant under an improper rotation $R(\pi/6) M_z$. 
The primitive lattice vectors of layer 1 are taken as  $\Vec{a}_1 = a(1,0)$ and $\Vec{a}_2 = a(1/2,\sqrt{3}/2)$,
where $a$ is the lattice constant,
and those of the layer 2 as $\tilde{\Vec{a}}_i = R(\pi/6)\,\Vec{a}_i$.
The atomic positions are given by Eq.\,\eqref{eq_atomic_position},
where $X=A,B$ ($\tilde{X}=\tilde{A},\tilde{B}$) denotes the sublattice site of layer 1 (2),
and $\GVec{\tau}_X$ and $\GVec{\tau}_{\tilde{X}}$ are the sublattice positions in the unit cell,
defined by $\GVec{\tau}_A=-\GVec{\tau}_1$, $\GVec{\tau}_B=\GVec{\tau}_1$,
$\GVec{\tau}_{\tilde{A}}= - R(\pi/6)\GVec{\tau}_1 + d\Vec{e}_z$, $\GVec{\tau}_{\tilde{B}}=R(\pi/6)\GVec{\tau}_1+ d\Vec{e}_z$
with $\GVec{\tau}_1 = (0,a/\sqrt{3})$,
where $d$ is the interlayer spacing between the two layers.
The reciprocal lattice vectors
of layer 1 are given by  $\Vec{a}^*_1 = (2\pi/a)(1,-1/\sqrt{3})$ and $\Vec{a}^*_2=(2\pi/a)(0,2/\sqrt{3})$,
and layer 2 by $\tilde{\Vec{a}}^*_i = R(\pi/6)\, \Vec{a}^*_i$.
The red and blue hexagons in
Fig.~\ref{fig_atomic_structure_and_BZ}(e) and (f)
show the Brillouin zones
of layer 1 and 2 in dodecagonal vdW-QCs, respectively.

\subsection{\label{sec:tb_model}Tight-binding model for van der Waals bilayers}

We model both systems by the tight-binding model with
spherical harmonic orbitals
of arbitrary atomic species.
Although we use a single $p_z$ orbital in this paper,
just like the model of graphene and hexagonal boron nitride,
but it can be any other orbital optimal for each system.
We discuss the effects of using other kinds of orbitals
in Sec.~\ref{sec:model_general_orbitals}.

The Hamiltonian is spanned by the Bloch bases
of each sublattice,
\begin{align}
	& |\Vec{k},X\rangle = 
	\frac{1}{\sqrt{N}}\sum_{\Vec{R}_{X}} e^{i\Vec{k}\cdot\Vec{R}_{X}}
	|\Vec{R}_{X} \rangle\quad (\mbox{layer 1}), \nonumber\\
	& |\tilde{\Vec{k}},\tilde{X}\rangle = 
	\frac{1}{\sqrt{N}}\sum_{\Vec{R}_{\tilde{X}}} e^{i\tilde{\Vec{k}}\cdot\Vec{R}_{\tilde{X}}}
	|\Vec{R}_{\tilde{X}}\rangle \quad (\mbox{layer 2}),
	\label{eq_bloch_base}
\end{align}
where $|\Vec{R}_{X} \rangle$ is the atomic orbital at the site $\Vec{R}_{X}$, 
$\Vec{k}$ and $\tilde{\Vec{k}}$ are the two-dimensional Bloch wave vectors
and $N = S_{\rm tot}/S$ is the number of the unit cells
with an area $S$
[$S=a^2$ for square lattices and $S=(\sqrt{3}/2)a^2$ for hexagonal lattices]
in the total system area $S_{\rm tot}$. 
We use a two-center Slater-Koster parametrization \cite{slater_koster,moon2012energy}
for the transfer integral between any two $p_z$ orbitals,
\begin{equation}
-T(\Vec{R}) = 
V_{pp\pi}\left[1-\left(\frac{\Vec{R}\cdot\Vec{e}_z}{|\Vec{R}|}\right)^2\right]
+ V_{pp\sigma}\left(\frac{\Vec{R}\cdot\Vec{e}_z}{|\Vec{R}|}\right)^2,
\label{eq_slater_koster}
\end{equation}
where
$\Vec{R}$ is the relative vector between two atoms, and
\begin{eqnarray}
V_{pp\pi} &&=  V_{pp\pi}^0 e^{- (|\Vec{R}|-a)/\delta_1}, \nonumber\\
V_{pp\sigma} &&=  V_{pp\sigma}^0  e^{- (|\Vec{R}|-3a)/\delta_2},    
\end{eqnarray}
for square lattices and
\begin{eqnarray}
V_{pp\pi} &&=  V_{pp\pi}^0 e^{- (|\Vec{R}|-a/\sqrt{3})/\delta_1}, \nonumber\\
V_{pp\sigma} &&=  V_{pp\sigma}^0  e^{- (|\Vec{R}|-1.36a)/\delta_2},    
\end{eqnarray}
for hexagonal lattices
so that the first-nearest neighbor intralayer coupling
becomes $V_{pp\pi}^0$.
In both systems, we take
$V_{pp\pi}^0 < 0$
and the decay length
of the transfer integral as
$\delta_1=\delta_2=0.184a$ \cite{TramblydeLaissardiere2010}.

 
The total tight-binding Hamiltonian is expressed as
\begin{equation}
\mathcal{H} = \mathcal{H}_{1} + \mathcal{H}_{2} + \mathcal{U} + \mathcal{H}_V + \mathcal{H}_{\Delta}
\label{eq_H_total}
\end{equation}
where $\mathcal{H}_{1}$ and $\mathcal{H}_{2}$ are the Hamiltonian for
the intrinsic square or hexagonal lattices of layer 1 and 2, respectively,
$\mathcal{U}$ is for the interlayer coupling,
$\mathcal{H}_V$ is for the interlayer potential asymmetry, and
$\mathcal{H}_{\Delta}$ is for the sublattice potential asymmetry.
The intralayer matrix elements of layer 1 are given by
\begin{align}
&\langle\Vec{k}',X' | \mathcal{H}_1 + \mathcal{H}_V + \mathcal{H}_\Delta |\Vec{k},X\rangle \nonumber\\
& =  \left[ h_{X,X'}(\Vec{k}) + \{V/2 + s_{\Delta} \Delta/2\} \delta_{X,X'} \right] \delta_{\Vec{k}', \Vec{k}},
\label{eq_H0}
\end{align}
where
\begin{equation}
h_{X,X'}(\Vec{k})  = \sum_{\Vec{L}}
-T(\Vec{L}+\GVec{\tau}_{X'X}) e^{-i\Vec{k}\cdot(\Vec{L}+\GVec{\tau}_{X'X})},
\end{equation}
and $\Vec{L} = n_1 \Vec{a}_1 + n_2 \Vec{a}_2$, $\GVec{\tau}_{X'X} = \GVec{\tau}_{X'}- \GVec{\tau}_{X}$,
$V$ is the magnitude of the interlayer potential asymmetry,
$s_{\Delta}$ is $+1$ and $-1$ for $X=A$ and $B$, respectively, and
$\Delta$ is the magnitude of the sublattice potential asymmetry.
Note that the square lattices with the minimal, one orbital model
considered in this work
does not have the $\mathcal{H}_{\Delta}$ term as there is only one sublattice in this case.
Similarly, the matrix for $H_2$ is given by replacing $\Vec{k}$ with $R(-\pi/4)\Vec{k}$ in a square lattice and
$R(-\pi/6)\Vec{k}$ in a hexagonal lattice,
$V/2$ with $-V/2$, and 
$s_{\Delta}$ by $+1$ and $-1$ for the sublattice
$\tilde{A}$ and $\tilde{B}$, respectively.
For $\mathcal{H}_1$ and $\mathcal{H}_2$,
we consider only the nearest neighbor interactions
in octagonal vdW-QCs
to keep the symmetric and simple cosine bands,
and all interactions of $|\Vec{R}|$ within $3a$
in dodecagonal vdW-QCs
to make it consistent with previous works \cite{Moon2013,Ahn2018,moon2019quasicrystalline}.

The interlayer matrix element between layer 1 and 2
is written as \cite{mele2010commensuration,bistritzer2011moire,koshino2015interlayer}
\begin{align}
& \langle\tilde{\Vec{k}},\tilde{X}| \mathcal{U}	|\Vec{k},X\rangle \nonumber\\
& = 
-\sum_{\Vec{G},\tilde{\Vec{G}}}
{t}(\Vec{k}+\Vec{G})
e^{-i\Vec{G}\cdot\mbox{\boldmath \scriptsize $\tau$}_{X}
	+i\tilde{\Vec{G}}\cdot\mbox{\boldmath \scriptsize $\tau$}_{\tilde{X}}}
\, \delta_{\Vec{k}+\Vec{G},\tilde{\Vec{k}}+\tilde{\Vec{G}}},
\label{eq_matrix_element_of_U}
\end{align}
where $\Vec{G}=m_1 \Vec{a}^*_1 + m_2 \Vec{a}^*_2$ and
$\tilde{\Vec{G}}=\tilde{m}_1 \tilde{\Vec{a}}^*_1 + \tilde{m}_2 \tilde{\Vec{a}}^*_2$
($m_1, m_2, \tilde{m}_1, \tilde{m}_2 \in \mathbb{Z}$)
run over all the reciprocal points
of layer 1 and 2, respectively.
Here
\begin{eqnarray}
{t}(\Vec{q}) = 
\frac{1}{S} \int
T(\Vec{r}+ z_{\tilde{X}X}\Vec{e}_z) 
e^{-i \Vec{q}\cdot \Vec{r}} d\Vec{r}
\label{eq_ft}
\end{eqnarray}
is the in-plane Fourier transform of the transfer integral,
where $z_{\tilde{X}X} = (\GVec{\tau}_{\tilde{X}}-\GVec{\tau}_{X})\cdot\Vec{e}_z$.
Note that both $T(\textbf{R})$ and $t(\textbf{q})$
between two $p_z$ orbitals
in the framework of a two-center Slater-Koster parametrization
are isotropic along the in-plane direction,
i.e., $T(\Vec{R})=T(|\Vec{R}|)$ and $t(\Vec{q})=t(|\Vec{q}|)$.

\subsection{Dual tight-binding model in momentum space}

In real-space, since quasicrystals do not have periodicity,
we need infinitely many atomic orbital bases to solve
Eq.~\eqref{eq_H_total}.
Although some conventional approximations
with a finite number of bases,
such as a periodic approximant
or a finite-size model,
can give an energy spectrum
quite similar to the actual spectrum,
the resulting wave functions
lose
their long-range quasicrystalline nature,
and spurious states, such as the boundary states, can
emerge.
In addition,
the use of the conventional theory on periodic moir\'{e} superlattices
cannot be validated in vdW-QCs
due to the absence of the moir\'{e} periodicity.

Instead, we can solve Eq.\,\eqref{eq_H_total} rigorously
by using a tight-binding model in momentum space,
which is the dual counterpart of the original
tight-binding Hamiltonian in the real space.
Equation (\ref{eq_matrix_element_of_U}) shows that the interlayer interaction occurs
between the states satisfying
the generalized Umklapp scattering condition $\Vec{k} + \Vec{G} = \tilde{\Vec{k}} + \tilde{\Vec{G}}$.
It is straightforward to show that
the entire subspace spanned by $\mathcal{H}_1 + \mathcal{H}_2 + \mathcal{U}$
from a layer 1's Bloch state at $\hat{\Vec{k}}$ is given by
$\{|\Vec{k},X\rangle \, | \, \Vec{k} = \hat{\Vec{k}} + \tilde{\Vec{G}}, \forall\tilde{\Vec{G}} \}$
and
$\{|\tilde{\Vec{k}},\tilde{X}\rangle \, | \, \tilde{\Vec{k}} = \hat{\Vec{k}} + \Vec{G}, \forall\Vec{G} \}$.
According to Eq.\ (\ref{eq_matrix_element_of_U}), the interaction strength between 
$\Vec{k}=\hat{\Vec{k}} + \tilde{\Vec{G}}$ and $\tilde{\Vec{k}}= \hat{\Vec{k}} + \Vec{G}$ is given by $t(\Vec{q})$ where 
$\Vec{q}=\Vec{k}+\Vec{G}= \tilde{\Vec{k}} + \tilde{\Vec{G}}=\Vec{k}-(\hat{\Vec{k}}-\tilde{\Vec{k}})$.
Then, the interaction strength can be visualized by
the diagram Figs.~\ref{fig_atomic_structure_and_BZ}(b) and (e), where
all the layer 2's wave points $\tilde{\Vec{k}}$ are inverted 
with respect to $\hat{\Vec{k}}$, i.e.,
$(\hat{\Vec{k}} - \tilde{\Vec{k}})$, and overlapped with the layer 1's wave points $\Vec{k}$.
In the map, the quantity $|\Vec{q}|=|\Vec{k}-(\hat{\Vec{k}}-\tilde{\Vec{k}})|$ is the geometrical distance between two points, 
and the interaction takes place only between the points located in close distance,
since $t(\Vec{q})$ decays in large $\Vec{q}$.
If the $k$ points are viewed as ``sites'', the whole system can be recognized as 
a tight-binding lattice in the momentum space,
which is dual to the original Hamiltonian in the real space.
This enables us to calculate
the electronic structures of almost every possible
stack of atomic layers
without
relying on moir\'{e} periodicity.

In this momentum-space tight-binding model,
the hopping between different momentum space sites
(the interlayer interaction $\mathcal{U}$)
is an order of magnitude smaller than the potential landscape (the band energies of the monolayers).
Thus, the eigenfunctions tend to be localized in momentum space
in a similar manner to the Aubry-Andr\'{e} model in one dimensional
real-space lattice under incommensurate perturbation \cite{aubry1980analyticity}.
Therefore, in practical calculation, we only need a limited number of states around $\hat{\Vec{k}}$ inside a certain cut-off circle $k_c$.
The $k_c$ should be greater than the typical localization length in momentum space,
but need not be too large, since the wave points discarded outside $k_c$
are properly accounted for by shifting $\hat{\Vec{k}}$
within the first Brillouin zone.
If we increase $k_c$, we will see more and more replicas of the identical quasi-energy band with different origins,
because shifting $\hat{\Vec{k}}$ actually corresponds to taking a different origin in the momentum space map of Figs.\,\ref{fig_atomic_structure_and_BZ}(b) and \ref{fig_atomic_structure_and_BZ}(e).
The analysis on the validity of the momentum-space cut-off
in Ref.~\cite{moon2019quasicrystalline} shows that
most states in van der Waals bilayers are made up of
20 or less monolayer states.
Thus, $k_c \approx 20/a$,
which includes a few hundreds of monolayer states within,
is sufficient in most practical calculation,
and we can obtain the energy eigenvalues at $\hat{\Vec{k}}$ by diagonalizing the Hamiltonian matrix within this finite set of bases.

\subsection{\label{sec:RND:resonant_condition}Resonant states respecting the rotational symmetry of quasicrystals}

The rotational symmetry of the quasicrystal
as well as the translational symmetries
of the constituent atomic layers
[Eq.~\eqref{eq_matrix_element_of_U}]
reveal the most dominant interaction,
which comes from the  resonance between degenerate states,
in each vdW-QC.
In octagonal vdW-QCs,
we see that the eight symmetric points 
$\Vec{Q}_n = [-\pi/a, \pi/a] + (a^*/\sqrt{2}) [\cos (n\pi/4), \sin (n\pi/4)]$
$(n=0,1,2,\cdots, 7)$
form a circular chain in the dual tight-binding lattice 
with $\hat{\Vec{k}}\,$=$\,\Vec{Q}_0$.
The chain has a radius of $a^*/\sqrt{2} \equiv |\Vec{a}^*_i|/\sqrt{2}=\sqrt{2}\pi/a$, 
and is indicated by the dashed ring in Fig.~\ref{fig_atomic_structure_and_BZ}(b).
Noting that the layer 2's wave points are inverted, 
these points are associated with layer 1's Bloch wave numbers $\Vec{k} = \Vec{Q}_n$ for even $n$'s
and layer 2's $\tilde{\Vec{k}} = \Vec{Q}_0 -\Vec{Q}_n$ for odd $n$'s.
Figure \ref{fig_atomic_structure_and_BZ}(c) shows 
the original positions of $\Vec{k}$ (layer 1) and $\tilde{\Vec{k}}$ (layer 2)
associated with $\Vec{Q}_n$ in the first Brillouin zone,
$\Vec{C}_n = \sqrt{2} a^* \sin (\pi/8) [\cos(3\pi/8+5n\pi/4), \sin(3\pi/8+5n\pi/4)]$
$(n=0,1,2,\cdots, 7)$.
Each intrinsic square lattice has a single cosinusoidal band
with a band maximum and minimum at 
$\mathrm{M}$- and $\Gamma$-points of the Brillouin zone, respectively.
Due to the symmetry, the  Bloch states of the intrinsic lattices
at the eights points are all degenerate in energy,
and therefore the interlayer coupling hybridizes them to make quasicrystalline resonant states.
Here the coupling is only relevant between the neighboring sites of the ring,
and it is given by $t_0 \equiv t(|\Vec{C}_n|)$.
The interaction to other neighboring states
in the dual tight-binding lattice
can be safely neglected 
since the interaction strength is much less than $t_0$ and
the two states are not degenerate in most cases.

Likewise, in dodecagonal vdW-QCs,
we see that the twelve symmetric points 
$\Vec{Q}_n = a^* [\cos (n\pi/6), \sin (n\pi /6)]$ $(n=0,1,2,\cdots, 11)$
form a circular chain in the dual tight-binding lattice
with $\hat{\Vec{k}}\,$=$\,\Vec{0}$ 
and the
radius is $a^*\equiv |\Vec{a}^*_i|=4\pi/(\sqrt{3}a)$
[dashed ring in Fig.~\ref{fig_atomic_structure_and_BZ}(e)].
These points are associated with layer 1's Bloch wave numbers $\Vec{k} = \Vec{Q}_n$ for even $n$'s
and layer 2's $\tilde{\Vec{k}} = -\Vec{Q}_n$ for odd $n$'s,
and Fig.~\ref{fig_atomic_structure_and_BZ}(f) shows 
the original positions of $\Vec{k}$ (layer 1)
and $\tilde{\Vec{k}}$ (layer 2) associated with $\Vec{Q}_n$
in the first Brillouin zone,
$\Vec{C}_n = 2a^*\sin (\pi/12) [\cos(5\pi/12+7n\pi/6), \sin(5\pi/12+7n\pi/6)]$
$(n=0,1,2,\cdots, 11)$.
Again, the Bloch states of each intrinsic hexagonal lattice at the twelve points are all degenerate in energy
and hybridized to form quasicrystalline resonant states by interlayer coupling
with $t_0 \equiv t(|\Vec{C}_n|)$.

It should be noted that these states are not the only set of states
which show the resonant coupling in each system.
As we shown in Appendix \ref{sec:App:weaker_resonant_condition},
there are more sets of states, with different wave numbers, that
show the resonant interaction
respecting the rotational symmetry of the quasicrystals.
However, the sets in Figs.~\ref{fig_atomic_structure_and_BZ}(b)
and (e) give the strongest interaction,
i.e., largest energy separation between the quasicrystalline states,
since these states form the rings with the shortest
distance between neighboring states
in the dual tight-binding lattices.



\subsection{\label{sec:model_general_orbitals}Quasicrystals from general atomic layers}

The quasicrystalline resonant states emerge from
the degeneracy of the constituent monolayer states
and the equivalence of the interlayer coupling strength
at all of the symmetric points $\textbf{C}_n$ in the Brillouin zone.
The former is always guaranteed
by the symmetry of atomic layers
as long as we use two identical atomic layers.
The latter, however, depends on the atomic orbitals involved.
In Sec.~\ref{sec:tb_model}, we described the Hamiltonian
of van der Waals bilayers
of which electronic structures are described mainly by
a single $p_z$ orbital at each atomic site.
However, we can obtain the quasicrystalline configuration from
any kind of atomic layers
with a square or hexagonal lattice symmetry,
some of which are
better described by, one or more, other types of orbital.


It is straightforward to show that
the transfer integral and interlayer coupling strength
between the atomic orbitals with 
spherical harmonics $Y_l^m$ and $Y_{l'}^{m'}$,
where $l$ is the angular momentum quantum numbers
and $m$ stands for the magnetic quantum number of atomic orbitals
(only in this subsection),
are
\begin{align}
-T(\textbf{R}) &= F(|\textbf{R}|) e^{i(m'-m)\phi_\textbf{R}}, \nonumber\\
-t(\textbf{q}) &= f(|\textbf{q}|) e^{i(m'-m)(\phi_\textbf{q}+\pi/2)},
\label{eq_tq_anisotropy}
\end{align}
where $\phi_\textbf{R}$ and $\phi_\textbf{q}$
are the azimuthal angle of $\textbf{R}$ and $\textbf{q}$
measured from $x$ axis to counterclockwise direction, respectively, and
$F(|\textbf{R}|)$ and $f(|\textbf{q}|)$ are real functions
that do not depend on $\phi_\textbf{R}$ and $\phi_\textbf{q}$, respectively.
Equation \eqref{eq_tq_anisotropy} shows that
$t(\Vec{q})$ between the atomic orbitals with the same $m$ is isotropic,
while that between different $m$ is not;
it is instead $|m-m'|$-fold rotational symmetric
and works as a potential with $1/|m-m'|$ period of the ring
in the dual tight-binding lattice.
Thus, replacing the $p_z$ orbital in Sec.~\ref{sec:tb_model}
with another spherical harmonic orbital
changes only the magnitude of quasicrystalline interaction
and does not influence the topology of the quasicrystalline bands
and the symmetry of the wave functions.
Meanwhile, vdW-QCs composed of atomic layers
with multiple atomic orbitals
exhibit the resonant interaction
respecting the rotational symmetry of the quasicrystals
if all the dominant interlayer interactions
occur between the atomic orbitals having the same $m$;
otherwise, they exhibit the resonant interaction
with a lower rotational symmetry.

In transition metal dichalcogenides (TMDC) monolayers,
both the first conduction band and valence band
are predominantly from
the $d_{z^2}$ and $d_{xy}+d_{x^2-y^2}$ (i.e., $Y_2^0$ and $Y_2^{\pm1}$) orbitals
of the metal atoms \cite{PhysRevB.84.153402,PhysRevB.88.085433,PhysRevB.92.205108}.
In TMDC bilayers, however,
the interaction between the $d$-orbitals in different layers
is negligible
since the metal atoms in different TMDC layers
are largely separated.
Instead, the dominant interlayer interaction comes from
the orbital hybridization
between the $p_z$ ($Y_1^0$) orbitals of the adjacent chalcogen layers
(of an order of sub-eV in $\mathrm{MoS}_2$ bilayer)
\cite{PhysRevB.88.075409,PhysRevB.92.205108,liu2015electronic},
and the next strongest interaction
(of an order of tens of meV in $\mathrm{MoS}_2$ 2H bilayer)
comes from
the coupling between $d_{z^2}$ and $p_z$ \cite{PhysRevB.92.205108}.
Thus, we need all of these orbitals, $Y_2^0$, $Y_2^{\pm1}$, $Y_1^0$,
to describe
the electronic structures of the quasicrystals composed of TMDC layers.
As Eq.~\eqref{eq_tq_anisotropy} shows,
the strongest (between $Y_1^0$)
and the next strongest (between $Y_2^0$ and $Y_1^0$) interlayer interaction
exhibit isotropic $t(\textbf{q})$.
Although $t(\textbf{q})$ between $Y_2^{\pm1}$ and $Y_1^0$ is anisotropic,
the interaction strength is at least 1 to 2 order weaker
than the strongest interaction
since the spatial extension of $Y_2^{\pm1}$ to the interlayer region
is less than that of $Y_2^0$.
Thus, most TMDC vdW-QCs will also show
the resonant states respecting the rotational symmetry of the quasicrystals.








\section{\label{sec:results_and_discussion}Results and discussion}

Below, we first investigate the electronic structures
of a vdW-QC
in the absence of the interlayer and sublattice potential asymmetry
by using a dual-tight binding method
with $t(\Vec{q})$
obtained at a specific combination of $(V_{pp\sigma}^0,d)$.
We investigate the electronic structures of vdW-QCs
predominantly described by a single spherical harmonic orbital
of any type.

Then we investigate
the change in the band structure
with respect to changes of constituent materials
or, equivalently, to changes in the interaction strength, which can be tuned by varying $t(\Vec{q})$.
If two different vdW-QCs
have the same sublattice configuration,
the interactions [Eq.\,\eqref{eq_matrix_element_of_U}]
in the two systems will have the same phase
and differ only in the magnitude of the interaction $t(\Vec{q})$.
Thus, the quasi-band dispersion near $\textbf{C}_n$
is mainly governed by the magnitude of $t_0=t(\Vec{C}_n)$,
together with the dispersion of the monolayer states,
and we can investigate the electronic structures of various vdW-QCs
by simply scaling $t_0$.
In addition, scaling $t_0$ also shows
the effects of tuning
the interlayer interaction
in a given vdW-QC with
the interlayer distance $d$,
e.g., by applying an external pressure or intercalation,
since the magnitude of the interlayer transfer integral
Eq.~\eqref{eq_slater_koster}
exponentially decays with $d$ \cite{koshino2015incommensurate}.

Then finally, we study the effects of breaking
the interlayer or sublattice symmetry.



\subsection{\label{sec:octa}Octagonal quasicrystal}

\subsubsection{Hamiltonian}

In octagonal vdW-QCs,
the strongest quasicrystalline resonant interaction occurs
at $\hat{\Vec{k}}=\Vec{Q}_0$.
By replacing $\hat{\Vec{k}}$
with $\Vec{Q}_0+\Vec{k}_0$,
we can express the Hamiltonian
\begin{equation}
    \mathcal{H} = \mathcal{H}_{\mathrm{ring}} + \mathcal{H}_V,
\label{eq_octa_H}
\end{equation}
in the vicinity of $\Vec{k}_0=\Vec{0}$,
in the bases of
$(|\Vec{k}^{(0)}\rangle,|\Vec{k}^{(1)}\rangle,\cdots,|\Vec{k}^{(7)}\rangle)$,
where $|\Vec{k}^{(n)}\rangle$ is
$|\Vec{k}_0+\Vec{C}_n,X\rangle$ for even $n$ (layer 1) and
$|\Vec{k}_0+\Vec{C}_n,\tilde{X}\rangle$ for odd $n$ (layer 2).
Here,
\begin{align}
&{\cal H}_{\rm ring}(\Vec{k}_0) = 
\begin{pmatrix}
H^{(0)} & -t_0 &&&& -t_0 \\
-t_0 & H^{(1)} & -t_0 \\
& -t_0 & H^{(2)} & -t_0 \\
&& \ddots & \ddots &\ddots \\
&&& -t_0 & H^{(6)} & -t_0 \\
-t_0 &&&& -t_0 & H^{(7)}
\end{pmatrix},
\label{eq_octa_H_ring}
\end{align}
is the Hamiltonian matrix of the resonant ring
in the absence of interlayer potential asymmetry,
where $H^{(n)}(\Vec{k}_0) = h_{X,X}[R(-5n\pi/4)\Vec{k}_0 + \Vec{C}_0]$, and
we neglect the $\Vec{k}_0$ dependence of the interlayer matrix element $t(\Vec{q})$.
The diagonal elements $H^{(n)}$ represent monolayer's Hamiltonian at $\Vec{k} = \Vec{k}_0 + \Vec{C}_n$ for even $n$
and $\tilde{\Vec{k}} =\Vec{k}_0 + \Vec{C}_n$ for odd $n$.
Note that $h_{X,X}$ in $H^{(n)}$ is same for any $n$, and
the dependence of the diagonal elements on $n$
solely comes from $R(-5n\pi/4) \Vec{k}_0$ in the argument of $h_{X,X}$.
Consequently, the ring Hamiltonian ${\cal H}_{\rm ring}$ is obviously symmetric under
rotation by a single span of the ring (i.e., moving $\Vec{C}_n$ to $\Vec{C}_{n+1}$),
which actually corresponds to the operation $[R(\pi/4) M_z]^5$ (225$^\circ$ rotation and swapping layer 1 and 2) in the original system.
In addition, $\mathcal{H}_{\mathrm{ring}}$ has a particle-hole symmetry
with respect to the energy $E=h_0$,
where $h_0 \equiv h_{X,X}(\Vec{C}_0) \approx -2\, V_{pp\pi}^0 (\cos\sqrt{2}\pi+1)$,
up to the first order to $\Vec{k}_0$ (Appendix \ref{sec:App:octa_k1st}).

$\mathcal{H}_V$ is the Hamiltonian representing
the interlayer potential asymmetry,
\begin{equation}
\mathcal{H}_V=\frac{V}{2}
\begin{pmatrix}
\sigma_z \\
& \sigma_z \\
&& \sigma_z \\
&&& \sigma_z
\end{pmatrix},
\end{equation}
where $V$ ($\ge 0$) represents the difference in the electrostatic energies
between the two layers,
and $\sigma_i$ is the Pauli matrix.
With $\mathcal{H}_V$,
the Hamiltonian $\mathcal{H}$,
which was originally in the form of one-dimensional monatomic chain
in the dual-tight binding lattice,
becomes that of diatomic chain with alternating on-site potential.

\subsubsection{\label{sec:octa_band_structure}Band structures and wave functions}

\begin{figure}[t]
	\begin{center}
		\leavevmode\includegraphics[width=1.0\hsize]{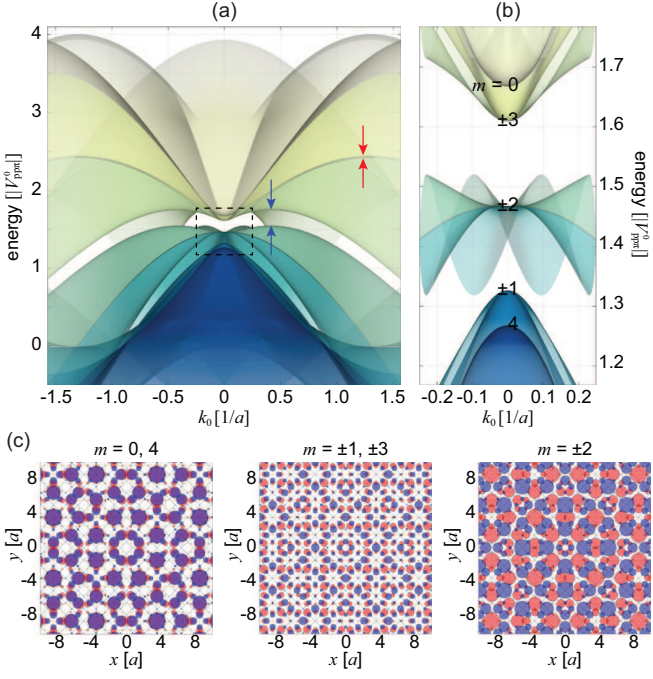}
	\end{center}
	\caption{
		(a) Electronic structure of octagonal vdW-QCs
		calculated by the 8-ring effective model.
		The blue and red arrows show the
		band opening by the interlayer
		and intralayer 2-wave mixing, respectively
		(see Sec.~\ref{sec:octa:DOS}).
(b) Detailed band structures near $\Vec{k}=\Vec{C}_n$
[the region encircled by the black dashed rectangle in (a)]
with index $m$ indicating the quantized angular momentum of the 8-fold rotational symmetry.
(c) LDOS at $\Vec{k}=\Vec{C}_n$ characterized by
$m$, where the area of the circle is proportional to the squared wave amplitude, and red and blue
circles represent the states in the upper and the lower layers, respectively.
		}
	\label{fig_octa_bs}
\end{figure}

\begin{figure}[t]
	\begin{center}
		\leavevmode\includegraphics[width=1.0\hsize]{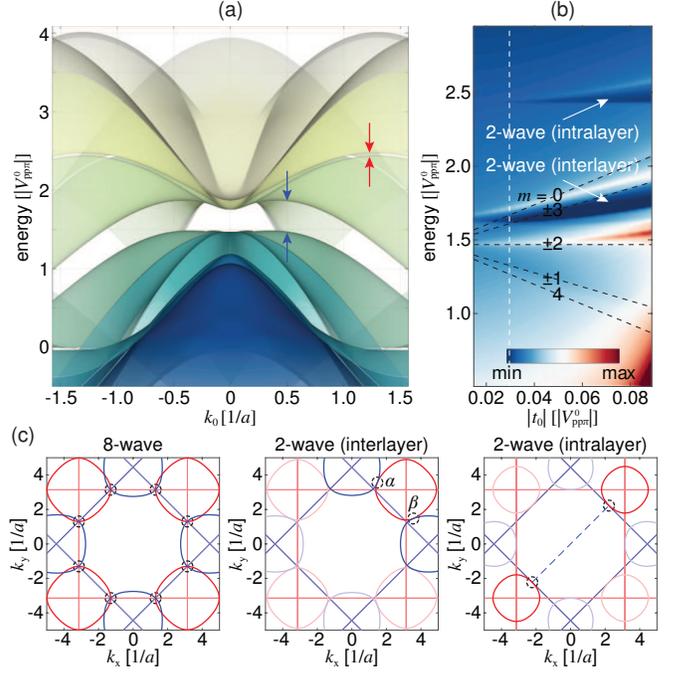}
	\end{center}
	\caption{
		(a) Electronic structures of octagonal vdW-QCs
		with the interlayer interaction $|t_0|$
		2 times larger than that in Fig.~\ref{fig_octa_bs}.
		The blue and red arrows show the
		band opening by the interlayer
		and intralayer 2-wave mixing, respectively.
		(b) Density map of DOS calculated by using
		32-waves
		and band edges of the quasicrystalline states [black dashed lines,
		Eq.\,\eqref{eq_octa_Em}]
		with various $t_0$.
		The white dashed line corresponds to the DOS
		for the system considered in Fig.~\ref{fig_octa_bs}.
		The white arrows show the pseudogaps opened
		by the interlayer and intralayer 2-wave interaction.
		(c) Three representative interactions
		residing in octagonal vdW-QCs;
		quasicrystalline interaction by 8-wave mixing (left),
		interlayer 2-wave mixing (middle), and
		intralayer 2-wave mixing (right).
		The red and blue contours show the Fermi surfaces
		of layer 1 and 2, respectively,
		the black dashed circles show the wave numbers
		where the interaction occurs, and
		the blue dashed line shows the reciprocal lattice vector
		of layer 2 which mediates the interaction
		between the states in layer 1.
	}
	\label{fig_octa_DOS_effects_of_t0}
\end{figure}

Figure \,\ref{fig_octa_bs}(a) shows
the band structures of the octagonal vdW-QCs near $\Vec{C}_n$,
in the absence of interlayer potential asymmetry (i.e., $V=0$),
plotted as a function of $\Vec{k}_\Vec{0}$,
and Fig.~\ref{fig_octa_bs}(b) shows
its closer view near $\Vec{k}_0=\Vec{0}$.
We choose
$(V_{pp\sigma}^0,d)=(-0.142\,V_{pp\pi}^0, 2.97a)$,
which gives
an interaction strength between the neighboring sites
of the circular chain in the dual-tight binding lattice of
$t_0=0.0296\,V_{pp\pi}^0$.
The eight parabolic bands are arranged on a circle with a radius
$\Delta k = (2-\sqrt{2})\pi/a$
by the Umklapp scattering [Eq.\,\eqref{eq_matrix_element_of_U}],
and they are strongly hybridized near $\Vec{k}_0=\Vec{0}$.
As a result,
the originally degenerate eight states of the square lattices
split into different energies and
exhibit characteristic dispersion,
including parabolic band-bottoms
and a frilled band edge,
which is flat up to the first order in $\Vec{k}_0$.

At $\Vec{k}_0=\Vec{0}$, $\mathcal{H}_{\mathrm{ring}}$
can be analytically diagonalized to obtain a set of energies
\begin{equation}
E_{m} =  h_0 - 2 t_0 \cos q_m,
\label{eq_octa_Em}
\end{equation}
which have the energy span of $4t_0$,
where $q_m = (5\pi/4) m$ with $m = 0, \pm1, \pm2, \pm3, 4$
the wave number along the chain.
Each element of the eigenvectors
$\Vec{v}_m=(\mu_m^{-3},\mu_m^{-2},\mu_m^{-1},\cdots,\mu_m^4)/\sqrt{8}$
($\mu_m=e^{i q_m}$)
is the coefficient to the Bloch bases
$|\Vec{k}^{(0)}\rangle,|\Vec{k}^{(1)}\rangle,\cdots,|\Vec{k}^{(7)}\rangle$.
Here the states with $m=\pm s$ $(s=1,2,3)$ form twofold doublets,
and belong to two-dimensional $E_s$ irreducible representation
of $D_{4d}$ point group,
while the $m=0$ and $4$ are non-degenerate, and belong to $A_1$ and $B_2$,
respectively.
If we disregard the $z$-position difference,
the index $m$ can be regarded as quantized angular momentum. The fact that there are 8 unique values for $m$
as well as the fact that
the eigenvalue of $R(\pi/4) M_z$ is given by $e^{i \pi m/4}$
are the evidence that the
quasicrystalline electronic states respect an 8-fold rotational symmetry.

The 8-wave resonant coupling also gives rise to a characteristic pattern in the wave function.
Figure\,\ref{fig_octa_bs}(c) shows the wave functions at $\Vec{k}_0=\Vec{0}$ where the hybridization is the most prominent. We can see that the wave amplitude is distributed on a limited number of sites in a 8-fold rotationally symmetric pattern.

\subsubsection{\label{sec:octa:DOS}The effect of the interlayer interaction and\\ 8- and 2-wave mixing}


As discussed in the beginning of this section,
we can calculate the quasicrystalline states
of various octagonal vdW-QCs,
which are either composed of other materials
or different interlayer distance $d$,
by simply scaling the magnitude of
the interlayer interaction
$t(\Vec{q})$.
Figure \ref{fig_octa_DOS_effects_of_t0}(a) shows
the band structures of an octagonal vdW-QC
with $t(\Vec{q})$
2 times larger than
that in Fig.~\ref{fig_octa_bs}.
Although the interaction strength varies with $\Vec{q}$,
hereafter we label each system
with $t(\Vec{q})$ at $\Vec{q}$
showing the strongest quasicrystalline interaction,
$t_0$ [$=t(\Vec{C}_n)$].
The stronger $t_0$ makes 
the 8-waves interact
over a much wider area in the Brillouin zone,
and the energy spacing between the quasicrystalline states larger.
Accordingly, the $m=\pm2$ states become flatter
and the band curvature of the other states increases.

The density map in Fig.~\ref{fig_octa_DOS_effects_of_t0}(b) shows
the density of states (DOS) of octagonal vdW-QCs
with various $t_0$,
with the white dashed line corresponding to the DOS
for the system considered in Fig.~\ref{fig_octa_bs}.
The large DOS observed at lower energies
in the systems with a large $|t_0|$
reflects the flat bands
arising from the 2-wave mixing
at $\Gamma$.
While the eight Bloch states centered at $\Vec{C}_n$
are sufficient to fully describe
the resonant interaction governing the quasicrystalline states,
some minor interactions at other wave vectors
are not captured by these bases.
Thus,
we used more (32 waves) bases
to calculate the DOS
in wider energy range.
We also plot the band edges of the quasicrystalline states
[Eq.\,\eqref{eq_octa_Em}]
by black dashed lines.
As $|t_0|$ increases,
the energy spacing between the edges increases
and the height of the DOS peaks also grows rapidly.
It should be noted that
some band edges (e.g., $m=\pm2$) lead to
a series of characteristic spiky peaks in DOS
and dips (pseudogaps) in between,
while other edges are buried
in the DOS of weakly coupled states.
Thus, quasicrystalline features,
such as local density of states (LDOS) with 8-fold rotational symmetry and relevant physical properties,
are most prominent at the energies
where the band edges coincide with the spiky peaks in DOS.
As changing $t_0$ does not break the symmetry of the Hamiltonian,
it neither changes the symmetry nor the degeneracy of quasicrystalline states.

In addition to the features from the quasicrystalline 8-wave mixing,
Fig.~\ref{fig_octa_DOS_effects_of_t0}(b)
shows the peaks and pseudogaps
associated with other kinds of interaction.
We plot the wave numbers
associated with these interactions
in Fig.~\ref{fig_octa_DOS_effects_of_t0}(c),
together with the Fermi surfaces.
The middle panel shows
the 2-wave mixing between the states in different layers \cite{Moon2013},
which occurs when the Fermi surfaces of
the two layers meet,
while the right panel shows
the 2-wave mixing between the states in the same layer
assisted by the potential of the opposite layer
\cite{koshino2015incommensurate,yao2018quasicrystalline}.
Blue and red arrows in
Figs.~\ref{fig_octa_bs}(a) and \ref{fig_octa_DOS_effects_of_t0}(a)
show the band opening
by the interlayer and intralayer 2-wave mixing, respectively,
whose size also increases with $|t_0|$.
The interlayer interaction strength $t(\Vec{q})$
involved in the interlayer and intralayer 2-wave mixing
is 0.472 and 1.49 times the interaction strength $t_0$
for the 8-wave interaction.
However, the intralayer mixing exhibits
a band opening smaller than the interlayer mixing
partly due to the two successive interlayer interaction
and partly due to the energy difference
between the states in opposite layers.
At $t_0 \rightarrow 0$ limit,
the 8-wave, interlayer 2-wave, intralayer 2-wave mixing
emerge at the energies
$E=\cos\sqrt{2}\pi+1 (\approx 1.47)$,
$-2\cos\frac{\sqrt{2}+1}{2}\pi (\approx 1.59)$,
$-4\cos(\pi/\sqrt{2}) (\approx 2.42)$
in unit of $|V_{pp\pi}^0|$.
It should be noted that the states and band opening arising from
these three mixing
are continuously connected to each other
in the Brillouin zone
[Figs.~\ref{fig_octa_DOS_effects_of_t0}(a) and (c)].
Unlike the quasicrystalline 8-wave interaction,
both the 2-wave mixing processes can occur in
bilayer square lattices stacked at any rotation angle $\theta$,
i.e., at usual moir\'{e} superlattices.
However,
it is straightforward to show that
$\alpha$ and $\beta$ in Fig.~\ref{fig_octa_DOS_effects_of_t0}(c),
which is typically known as a \textit{moir\'{e} interaction},
occur at different energies \cite{Moon2013}
in the systems with $\theta$ other than $45^\circ$.

\subsubsection{Effects of interlayer potential asymmetry}

\begin{figure}
	\begin{center}
		\leavevmode\includegraphics[width=1.0\hsize]{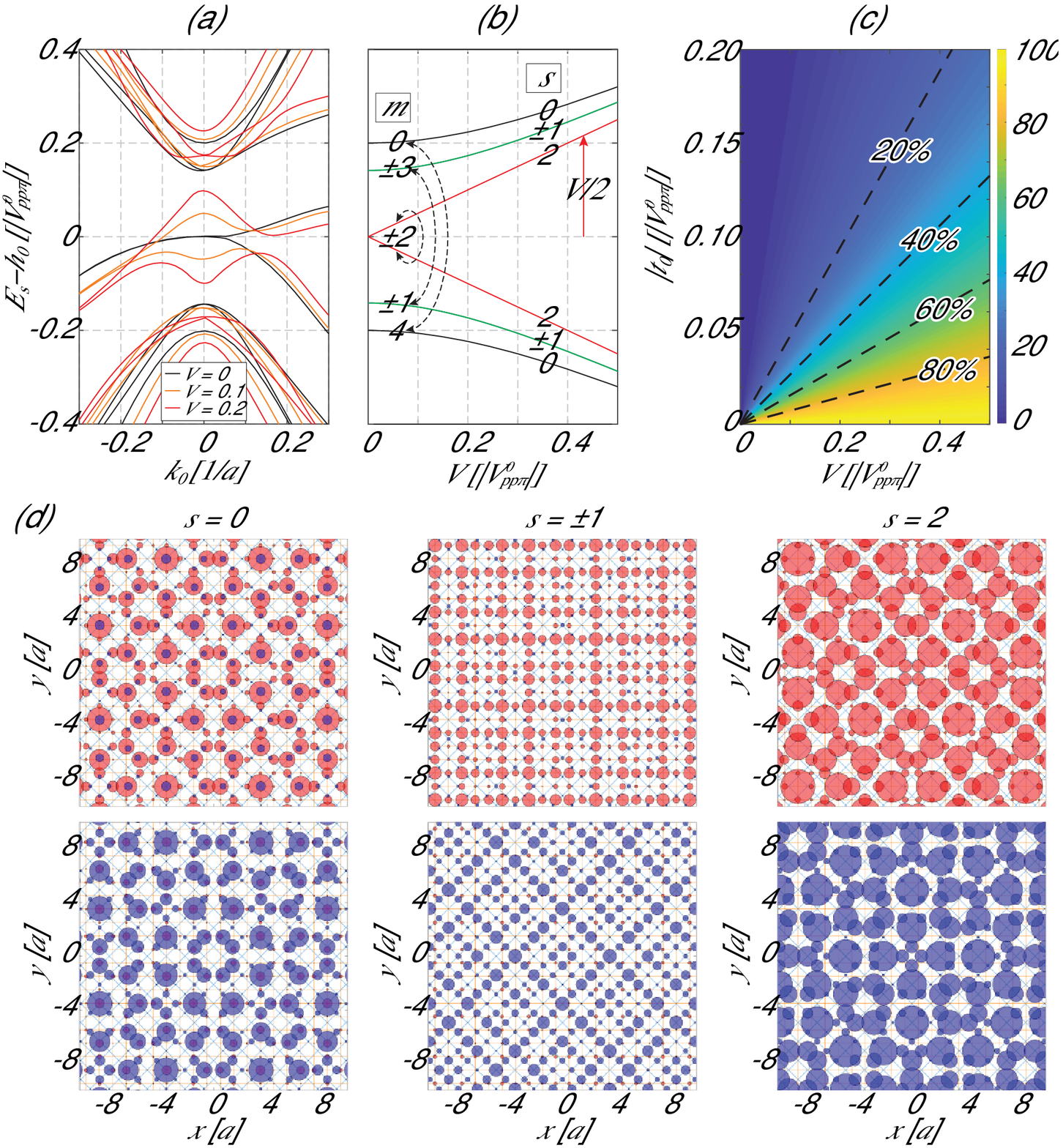}
	\end{center}
	\caption{
		(a) Band dispersion near $\Vec{C}_n$
		of octagonal vdW-QCs under three different interlayer potential asymmetry, $V=0$, $0.1\,|V_{pp\pi}^0|$, $0.2\,|V_{pp\pi}^0|$.
		(b) Band edges at $\Vec{C}_n$ with
		various $V$.
		Indices $m$ and $s$ show
		the angular momentum of the pristine quasicrystalline states
		with 8-fold rotational symmetry
		and that of 4-fold rotational symmetry
		under interlayer potential asymmetry.
		Dashed arrows show the interaction
		between the constituent quasicrystalline states by $\mathcal{H}_V$.
		(c) Degree of mixing (in percentage) between the constituent states
		in $\Psi_{s=\pm1}$ (see text)
		with respect to the interlayer interaction strength $|t_0|$ and
		interlayer potential asymmetry $V$.
		(d) Plots similar to Fig.~\ref{fig_octa_bs}(c)
		for $V\ne0$. The top and bottom panels show the LDOS
		of the upper and lower bands, respectively.
		}
	\label{fig_octa_bs_w_U}
\end{figure}

Figure \ref{fig_octa_bs_w_U}(a) shows the band dispersion
near $\Vec{C}_n$ of octagonal vdW-QCs
under three different interlayer asymmetric potential,
$V=0$, $0.1$, $0.2$ in unit of $|V_{pp\pi}^0|$.
Since Eq.\,\eqref{eq_octa_H} satisfies
$\Sigma'^{-1} (\mathcal{H}-h_0\mathbb{I}) \Sigma' = -(\mathcal{H}-h_0\mathbb{I})$
at $\Vec{k}_0=\Vec{0}$
for $\Sigma'=\mathrm{diag}(i\sigma_y,i\sigma_y,i\sigma_y,i\sigma_y)$
and
$\mathbb{I}$ is an $8\times8$ unit matrix,
regardless of the presence of the interlayer potential asymmetry,
$\mathcal{H}$ has a particle-hole symmetry
with respect to the energy $E=h_0$.
As $V$ increases, however, the states with $m=\pm2$ at $\Vec{k}_0=\Vec{0}$
lose the degeneracy,
and all the band edges
move away from $E_{m=\pm2}$ ($=h_0$).

We can obtain further insight
on the effects of
the interlayer potential asymmetry
from the analytic expression of the energies at $\Vec{k}_0=\Vec{0}$.
The interlayer potential asymmetry
couples the eigenstates of
$\mathcal{H}_{\mathrm{ring}}(\Vec{k}_0=\Vec{0})$
that have angular momenta that differ by $\pm4$,
\begin{equation}
\langle \Vec{v}_{m'} | \mathcal{H}_V | \Vec{v}_m \rangle =
\left\{ 
\begin{array}{cc}
 -V/2 & (m-m' \equiv 4\,(\mathrm{mod}\,8)), \\
 0 & (\mathrm{otherwise}),
\end{array}
\right.
\end{equation}
since the diagonal elements of $\mathcal{H}_V$
work as a staggered potential
with 1/4 period of the ring
in the dual tight-binding lattice.
Thus, the Hamiltonian matrix
in the bases of the quasicrystalline states is reduced to
four $2\times 2$ matrices
\begin{equation}
\tilde{\mathcal{H}}_{m,m'}=
\begin{pmatrix}
E_m & -V/2\\
-V/2 & E_{m'}
\end{pmatrix},
\label{eq_octa_H_E_of_U}
\end{equation}
for $(m,m')=(0,4),(1,-3),(2,-2),(3,-1)$.
As the quasicrystalline states, $\Vec{v}_m$,
originate from the resonant interaction
between the degenerate states $|\Vec{k}^{(n)}\rangle$
in the two layers,
the interlayer potential asymmetry breaks
the 8-fold rotational symmetry of the states
by lifting the degeneracy of $|\Vec{k}^{(n)}\rangle$. This reduces the
allowed
angular quantum numbers to  
$s\,(=0,\pm1,2)\equiv m \equiv m'\,(\mathrm{mod}\,4)$, which indicates a 4-fold rotational symmetry.
We obtain the following energies and wave functions
\begin{eqnarray}
E_s && =h_0 \pm \sqrt{4t_0^2 \cos^2(5\pi s/4) + V^2/4 }, \nonumber\\
\Psi_s && =c_m \Vec{v}_m + c_{m'} \Vec{v}_{m'},
\label{eq_octa_E_of_U}
\end{eqnarray}
where $(c_m, c_{m'})$ is
$(\sin(\phi/2), -\cos(\phi/2))$ for the upper band and
$(\cos(\phi/2), \sin(\phi/2))$ for the lower band
with $\phi = \tan^{-1}(V/(4t_0\cos q_m))$,
and plot $E_s$ against $V$
in Fig.~\ref{fig_octa_bs_w_U}(b).
The states with $s=1$ and $s=-1$ are always degenerate
due to the $\Vec{v}_m^* = \Vec{v}_{-m}$ symmetry of the wave functions.
At small $V$,
the interlayer interaction $|t_0|$
suppresses the energy shift of $s=0,\pm1$ states
[Eq.\,\eqref{eq_octa_E_of_U}],
in a similar manner to the way the interaction
suppresses the Dirac point shift in twisted bilayer graphene with a small twist angle
\cite{moon2014optical}.
On the other hand, the two states with $s=2$ $(m=\pm2)$
are composed of two degenerate quasicrystalline states,
$m=2$ and $m=-2$.
Thus, their band edges
shift as much as the applied bias
in opposite directions
and are not affected by the interlayer interaction $t_0$
as can be clearly seen from Eq.~\eqref{eq_octa_H_E_of_U}.
As $V$ increases,
the overall energy span of these resonant states increases,
while the energy spacing between
the adjacent states decreases.

The dashed arrows in Fig.~\ref{fig_octa_bs_w_U}(b) show the interaction
between quasicrystalline states between $\Vec{v}_m$ and $\Vec{v}_{m'}$
by $\mathcal{H}_V$,
and Fig.~\ref{fig_octa_bs_w_U}(c) shows
the degree of mixing in $\Psi_{s=\pm1}$,
which we defined as
$(1-||c_m|^2-|c_{m'}|^2|)\times100\,[\%]$.
Systems with $|t_0|<V$ exhibit
stronger mixing,
which will
influence the transition behavior,
such as the optical selection rule, in vdW-QCs.
The states with $s=0$ exhibit
a similar, but slightly weaker, mixing
owing to the larger energy difference between $E_m$ and $E_{m'}$
in $\Psi_{s=0}$.
However, the states with $s=2$ are special in that
the constituent states $\Vec{v}_2$ and $\Vec{v}_{-2}$
are always fully mixed,
i.e., $c_{m'}=-c_m$ for the upper band
and $c_{m'}=c_m$ for the lower band,
regardless of the values of $t_0$ and $V$. Again, this is
due to the degeneracy between
the constituent states $m=2$ and $m'=-2$.

We plot the LDOS of the upper and lower bands
with $s=0,\pm1,2$ at $V=0.2\,|V_{pp\pi}^0|$
in the top and bottom panels
in Fig.~\ref{fig_octa_bs_w_U}(d), respectively.
Due to the interlayer potential asymmetry,
the wave functions $\Psi_s$ are
more or less spatially polarized to either layer.
And the stronger the mixing,
the more the wave functions are layer polarized;
for example, $\Psi_{s=\pm1}$ exhibit
more polarization than $\Psi_{s=0}$.
This is because
\begin{equation}
\Psi_s=\frac{1}{2}(c_m-c_{m'})(\Vec{v}_m-\Vec{v}_{m'})
+\frac{1}{2}(c_m+c_{m'})(\Vec{v}_m+\Vec{v}_{m'}),
\label{eq_layer_polarization}
\end{equation}
where $(c_m, c_{m'}) \in \mathbb{R}$, and
$\Vec{v}_m-\Vec{v}_{m'}$ and $\Vec{v}_m+\Vec{v}_{m'}$ are
perfectly polarized to layer 1 and 2, respectively,
since $\mu_{m'}=-\mu_m$.
Thus, as the mixing becomes stronger,
the upper bands ($c_{m'} \approx -c_m$) consist mostly of
$|\Vec{k}^{(n)}\rangle$ with even $n$ (i.e., layer 1)
while the lower bands ($c_{m'} \approx c_m$) consist mostly of
$|\Vec{k}^{(n)}\rangle$ with odd $n$ (i.e., layer 2).
Again, the states with $s=2$ are special in that
their wave functions $\Psi_{s=2}$ are perfectly
polarized to either layer
regardless of the values of $t_0$ and $V$ because the constituent states $\Vec{v}_2$ and $\Vec{v}_{-2}$
are always fully mixed. This is similar to the case of an one-dimensional diatomic chain whose sublattices
stop completely at the acoustic and optical modes.

$\Vec{C}_n$ in Fig.~\ref{fig_atomic_structure_and_BZ}(c),
the wave vectors where quasicrystalline interaction occurs,
remain the same
since the interlayer potential asymmetry
does not change the Umklapp scattering paths.
Thus, the LDOS profile of each layer-polarized state,
which is associated with $\Vec{C}_n$ for even $n$ (layer 1) or
$\Vec{C}_n$ for odd $n$ (layer 2),
is exactly consistent with the profile of each layer
in the absence of the potential asymmetry
[Fig.~\ref{fig_octa_bs}(c)].
Therefore, we can switch
between the quasicrystalline states
and their layer components by applying an electric field.

\subsection{\label{sec:dodeca}Dodecagonal quasicrystal}

\subsubsection{Hamiltonian}

In dodecagonal vdW-QCs, the strongest quasicrystalline resonant interaction
occurs at $\hat{\Vec{k}}=\Vec{0}$.
Thus, by replacing $\hat{\Vec{k}}$ with $\Vec{k}_0$,
we can express the Hamiltonian of the resonant ring $\mathcal{H}_{\mathrm{ring}}$
in the absence of the interlayer and sublattice potential asymmetry
by a $24\times 24$ matrix
\begin{align}
&{\cal H}_{\rm ring}(\Vec{k}_0) = 
\begin{pmatrix}
H^{(0)} & W^\dagger &&&& W \\
W & H^{(1)} & W^\dagger \\
& W & H^{(2)} & W^\dagger \\
&& \ddots & \ddots &\ddots \\
&&& W & H^{(10)} & W^\dagger \\
W^\dagger &&&& W & H^{(11)}
\end{pmatrix},
\label{eq_dodeca_H_ring}
\\
&  H^{(n)}(\Vec{k}_0) = 
\begin{pmatrix}
h_{AA}^{(n)} & h_{AB}^{(n)}  \\
h_{BA}^{(n)} & h_{BB}^{(n)}  \\
\end{pmatrix},
\quad
W =  - t_0
\begin{pmatrix}
\omega & 1\\
1 & \omega^*
\end{pmatrix},
\end{align}
in the bases of
$(|\Vec{k}^{(0)}\rangle,|\tilde{\Vec{k}}^{(1)}\rangle,|\Vec{k}^{(2)}\rangle,|\tilde{\Vec{k}}^{(3)}\rangle,\cdots,|\Vec{k}^{(11)}\rangle)$.
Here, $\Vec{k}^{(n)}=\Vec{k}_0+\Vec{Q}_n$ for even $n$ (layer 1)
and $\tilde{\Vec{k}}^{(n)}=\Vec{k}_0-\Vec{Q}_n$ for odd $n$ (layer 2),
where $|\Vec{k}^{(n)}\rangle$ and $|\tilde{\Vec{k}}^{(n)}\rangle$ are
$(|\Vec{k}^{(n)},X\rangle,|\Vec{k}^{(n)},X'\rangle)$ and
$(|\tilde{\Vec{k}}^{(n)},\tilde{X}\rangle,|\tilde{\Vec{k}}^{(n)},\tilde{X}'\rangle)$
with the sublattices $X$ and $X'$ are arranged in the order of
$(A,B)$ or $(\tilde{A},\tilde{B})$
for $n\equiv0,3$ modulo 4,
and $(B,A)$ or $(\tilde{B},\tilde{A})$
for $n\equiv1,2$.
And $h_{X'X}^{(n)}(\Vec{k}_0) = h_{X'X}[R(-7n\pi/6) \Vec{k}_0 + \Vec{Q}_0]$, $\omega=e^{2\pi i/3}$,
and we neglect the $\Vec{k}_0$ dependence of the interlayer matrix element $t(\Vec{q})$.

\begin{figure}[t]
	\begin{center}
		\leavevmode\includegraphics[width=1.0\hsize]{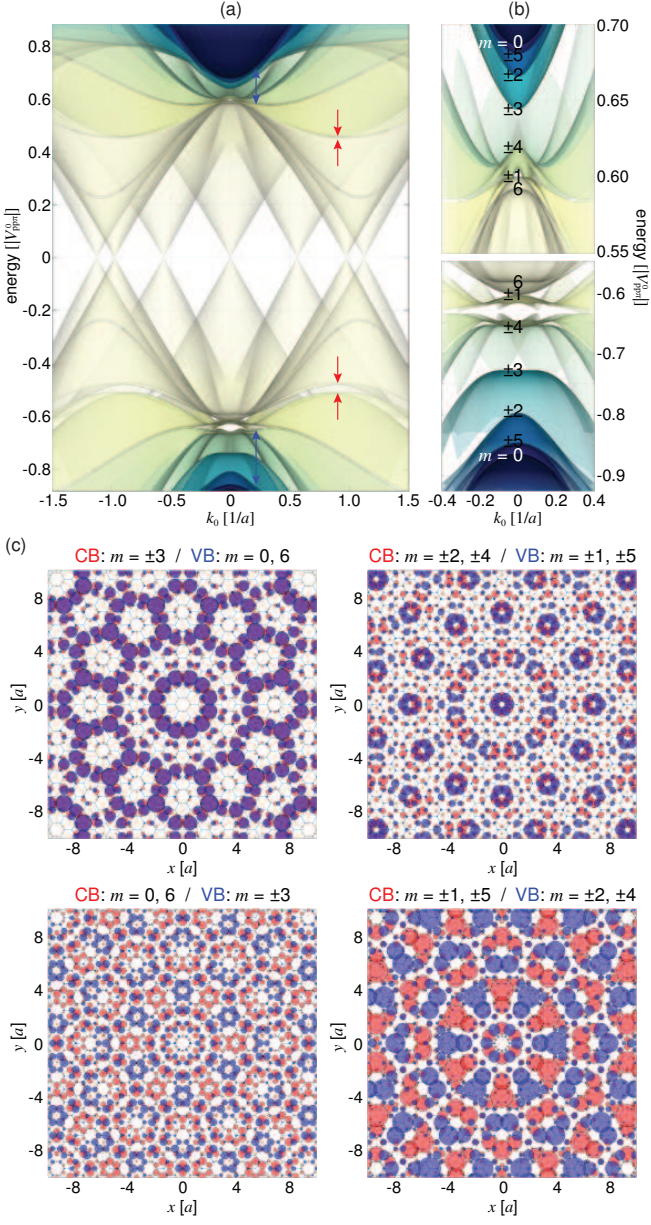}
	\end{center}
	\caption{
		Plots similar to Fig.~\ref{fig_octa_bs}
		for dodecagonal vdW-QCs
		calculated by the 12-ring effective model.
		The blue and red arrows in (a) show the
		band opening by the interlayer
		and intralayer 2-wave mixing, respectively.
		The top and bottom panels in (b)
		show the quasicrystalline states
		in the conduction band and valence band, respectively.
		}
	\label{fig_dodeca_bs}
\end{figure}

\begin{figure}[t]
	\begin{center}
		\leavevmode\includegraphics[width=1.0\hsize]{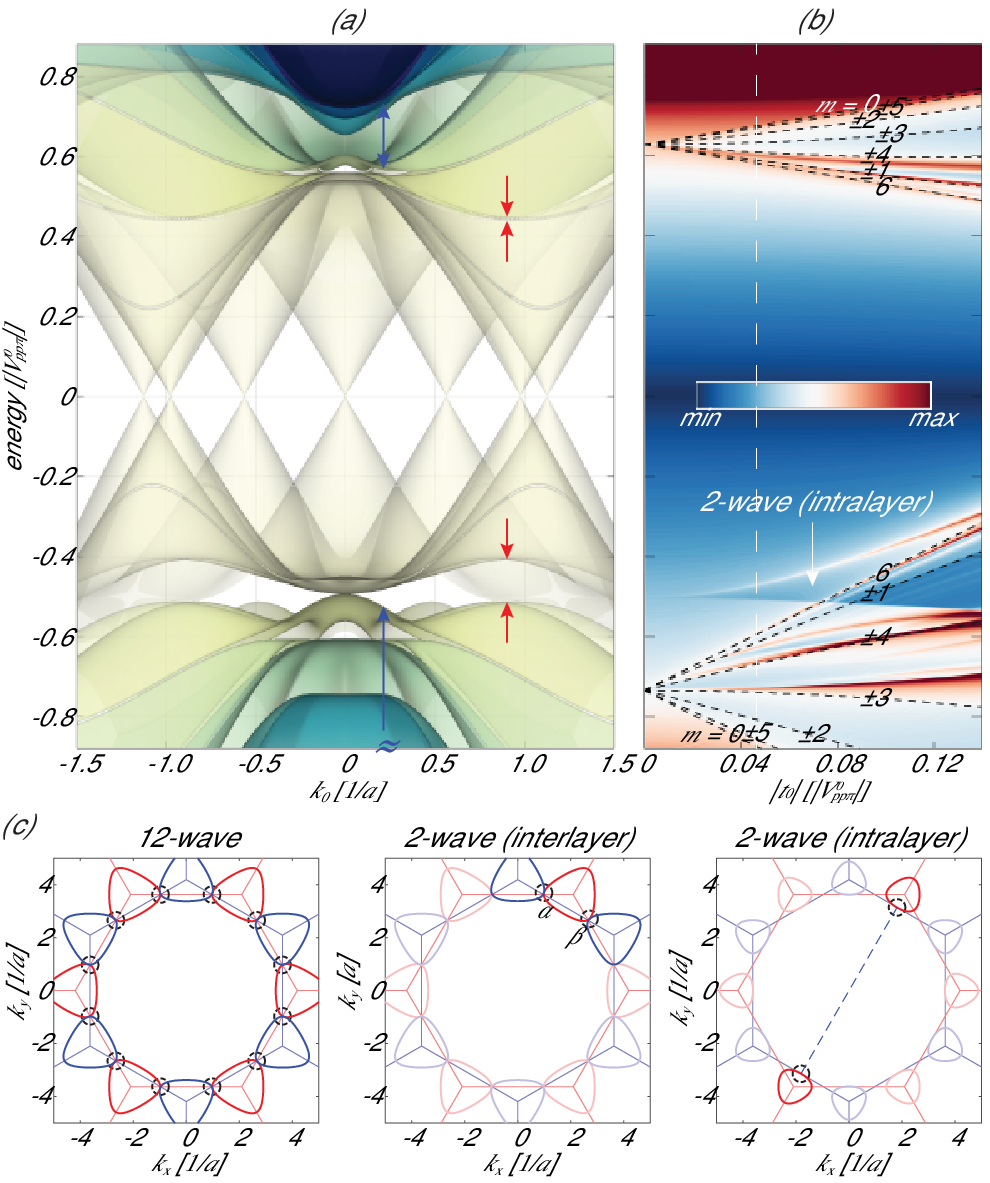}
	\end{center}
	\caption{
		(a) Electronic structures of dodecagonal vdW-QCs
		with the interlayer interaction $|t_0|$
		2 times larger than that in Fig.~\ref{fig_dodeca_bs}.
		The blue and red arrows show the
		band opening by the interlayer
		and intralayer 2-wave mixing, respectively.
		(b) Density map of DOS calculated by using 182-waves
		and band edges of the quasicrystalline states [black dashed lines,
		Eq.\,\eqref{eq_dodeca_Em}]
		with various $t_0$.
		The white dashed line corresponds to the DOS
		for the system considered in Fig.~\ref{fig_dodeca_bs}.
		(c) Plots similar to Fig.~\ref{fig_octa_DOS_effects_of_t0}(c)
		in dodecagonal vdW-QCs.
		}
	\label{fig_dodeca_DOS_effects_of_t0}
\end{figure}

In the given bases order,
the Hamiltonian representing
the interlayer and sublattice potential asymmetry are expressed by
\begin{equation}
\mathcal{H}_V=\frac{V}{2}\mathrm{diag}(
\mathbb{I},-\mathbb{I},
\mathbb{I},-\mathbb{I},
\mathbb{I},-\mathbb{I},
\mathbb{I},-\mathbb{I},
\mathbb{I},-\mathbb{I},
\mathbb{I},-\mathbb{I}
),
\end{equation}
and
\begin{eqnarray}
\mathcal{H}_\Delta=\frac{\Delta}{2}\mathrm{diag}(
&&\sigma_z,-\sigma_z,-\sigma_z,\sigma_z,
\sigma_z,-\sigma_z,-\sigma_z,\sigma_z, \nonumber\\
&&\sigma_z,-\sigma_z,-\sigma_z,\sigma_z
),
\end{eqnarray}
respectively,
where $V$ represents the difference in the electrostatic energies
between the two layers,
$\Delta$ is the difference between the on-site potentials
between two sublattices, and
$\mathbb{I}$ is a $2\times2$ unit matrix.
Then, the Hamiltonian of general dodecagonal vdW-QCs is given by
\begin{equation}
    \mathcal{H} = \mathcal{H}_{\mathrm{ring}} + \mathcal{H}_V + \mathcal{H}_\Delta.
\label{eq_dodeca_H}
\end{equation}

\subsubsection{\label{sec:dodeca_band_structure}Band structures and wave functions}

We plot the band structures near $\Vec{C}_n$ of the dodecagonal vdW-QCs with
$(V_{pp\sigma}^0,d)=(-0.142\,V_{pp\pi}^0, 1.36a)$ \cite{comment_on_parameters}
and $V=0$, $\Delta=0$ in Fig.~\ref{fig_dodeca_bs}(a),
and their closer view in (b).
The twelve Dirac cones are arranged on a circle with a radius
$\Delta k = 4(2-\sqrt{3})\pi/(3a)$,
and they are strongly hybridized near $\Vec{k}_0=\Vec{0}$
with $t_0=0.0465\,V_{pp\pi}^0$
to exhibit the characteristic dispersion
including flat band bottoms, the Mexican-hat edges,
and the frilled band edges.
We can get the electronic structures of various dodecagonal vdW-QCs
by using the proper $V_{pp\pi}^0$;
e.g., $V_{pp\pi}^0=-3.38\,\mathrm{eV}$ \cite{comment_on_parameters}
gives the spectrum of vdW-QC composed of two graphene layers,
which is known as quasicrystalline twisted bilayer graphene.

At $\Vec{k}_0=\Vec{0}$, $\mathcal{H}_{\mathrm{ring}}$
can be analytically diagonalized to obtain a set of energies (neglecting the constant energy)
\begin{align}
& E^\pm_{m} =  t_0 \cos q_m \pm \sqrt{3t_0^2 \sin^2 q_m + (h_0 - 2 t_0 \cos q_m)^2},
\label{eq_dodeca_Em}
\end{align}
where $h_0 \equiv h_{AB}(\Vec{Q}_0) = h_{BA}(\Vec{Q}_0) = -0.682\, V_{pp\pi}^0$,
and $q_m = (7\pi/6) m$ with $m = -5,-4,\cdots, 5,6$
is the wave number along the chain.
Unlike the octagonal vdW-QCs in the minimal model (Sec.\,\ref{sec:octa}),
which has one set of the hybridized states,
the dodecagonal vdW-QCs show hybridization
both in the conduction band and valence bands,
which correspond to $\pm$ in Eq.\,\eqref{eq_dodeca_Em}, respectively.
The energy scaling in the conduction band is, however,
much smaller than that in the valence band
since the wave function of the conduction band of
the hexagonal lattices,
having the same phases between the sublattices,
suppresses the interlayer interaction
by a factor of 3.
The index $m$ is a quantized angular momentum
respecting the 12-fold rotational symmetry.
The states with $m=\pm s$ $(s=1,2,3,4,5)$ form twofold doublets, while
the $m=0$ and $6$ are non-degenerate.
Note that the interaction responsible
for the formation of the quasicrystalline states
only weakly affects the spectrum
at energies away from the momentum matching conditions;
e.g., in a quasicrystalline twisted bilayer graphene
there is no meaningful change on the Fermi velocity \cite{Ahn2018},
since $E_m^\pm$ are far from the Dirac point.

Figure \ref{fig_dodeca_bs}(c) shows
the LDOS of the quasicrystalline states,
where we can see that the wave amplitude distribute selectively on a limited number of sites in a characteristic 12-fold rotationally symmetric pattern.
The wave functions for $E_m^\pm$ are
$\Vec{v}_m^\pm=(1/\sqrt{12})(\mu_m^{-5},\mu_m^{-4},\mu_m^{-3},\cdots,\mu_m^6)
\bigotimes (c_{m,1}^\pm,c_{m,2}^\pm)$
($\mu_m=e^{i q_m}$), where
$(c_{m,1}^+,c_{m,2}^+)=(\sin(\phi_m/2),\cos(\phi_m/2))$ and
$(c_{m,1}^-,c_{m,2}^-)=(\cos(\phi_m/2),-\sin(\phi_m/2))$ 
are the coefficients of the sublattices
arranged in the order of the bases of Eq.~\eqref{eq_dodeca_H_ring}, and
$\phi_m=\tan^{-1}[(h_0-2t_0 \cos q_m)/(\sqrt{3} t_0 \sin q_m)]$.
Since the Hamiltonian has a symmetry of
\begin{eqnarray}
&&\Sigma'^{-1} \mathcal{H}_{\mathrm{ring}} \Sigma
= \mathcal{H}_{\mathrm{ring}}^* \nonumber\\
&&\Sigma=\mathrm{diag}(\sigma_x,\sigma_x,\sigma_x,\sigma_x,\sigma_x,\sigma_x)
\end{eqnarray}
at $\Vec{k}_0=\Vec{0}$,
the states with angular momentum $m$ and $-m$ are degenerate
and $\Vec{v}_{-m}^\pm = \sigma_x (\Vec{v}_m^\pm)^*$,
and it is straightforward to show that
their LDOS profiles are exactly the same to each other.
Figure \ref{fig_dodeca_bs}(c)
also shows that the states with $\pm m$
exhibit LDOS profiles which look similar to those of $6\mp m$;
the analysis on the wave functions $\Vec{v}_m^\pm$ clearly shows that
the states with $m=0$ and $6$, and also the states with $m=3$ and $-3$
have LDOS that are exactly the same as each other,
while the LDOS profiles of the other states
(i.e., $m=\pm1$ and $\pm5$, and also $m=\pm2$ and $\pm4$)
become different
as $|t_0|$ grows.
Likewise, the $\pm m$ states in the conduction band
exhibit LDOS profiles which look similar to
the $3 \mp m$ ones in the valence band
in the systems with
a small $|t_0/V_{pp\pi}^0|$.

\subsubsection{The effect of the interlayer interaction and\\ 12- and 2-wave mixing}


Figure \ref{fig_dodeca_DOS_effects_of_t0}(a) shows
the valence band structures of a dodecagonal vdW-QC
with a interlayer interaction $t(\Vec{q})$ that is
2 times larger than
the one in Fig.~\ref{fig_dodeca_bs}.
The energy spacing between the quasicrystalline states
becomes larger, and
the flat band area of $m = 6$ $(m = \pm4)$ state
in the valence band
in Fig.~\ref{fig_dodeca_DOS_effects_of_t0}(a)
is approximately 2-times (5-times) as large as 
that in Fig.~\ref{fig_dodeca_bs}(a),
and it is 28-times (70-times) bigger than the flat band area
of magic-angle twisted bilayer graphene.
As a greater number of the electronic states are involved in the flat bands,
we expect to see stronger electron-electron interacting effect.
The density map in Fig.~\ref{fig_dodeca_DOS_effects_of_t0}(b) shows
the DOS of dodecagonal vdW-QCs with various $t_0$
calculated by using 182-wave bases.
The white dashed line corresponds to the DOS
for the system considered in Fig.~\ref{fig_dodeca_bs},
and the black dashed lines show
the band edges of the quasicrystalline states
[Eq.\,\eqref{eq_dodeca_Em}].

The systems with larger $|t_0|$
exhibit higher DOS peaks
owing to the increase of the flat band area in the momentum space.
Not every quasicrystalline state leads to a DOS displaying spiky peaks interspersed with pseudogaps,
so quasicrystalline features would be most prominent
at the energies
where the band edges coincide with the spiky peaks in DOS,
especially at the $m=6,\pm4,\pm3$ states in the valence band
and the $m=6,\pm1,\pm4$ states in the conduction band.
Again, changing $t_0$ neither changes the symmetry
nor degeneracy of quasicrystalline states.
In most practical parameter ranges,
a system with larger $|t_0|$ exhibits
a larger energy spacing between the quasicrystalline states
in both the conduction and valence bands.
Note that, in the systems with extremely strong interlayer interaction
($|t_0| > h_0/2$),
the energy spacing in the conduction band may decrease as $|t_0|$ increases
(Appendix \ref{sec:App:dodeca_band_edges}).
Such a condition, however, is hard to be achieved in the most practical systems.
Thus, hereafter, we will consider the systems with $|t_0| < h_0/2$ only.

\begin{figure*}
	\begin{center}
		\leavevmode\includegraphics[width=0.9\hsize]{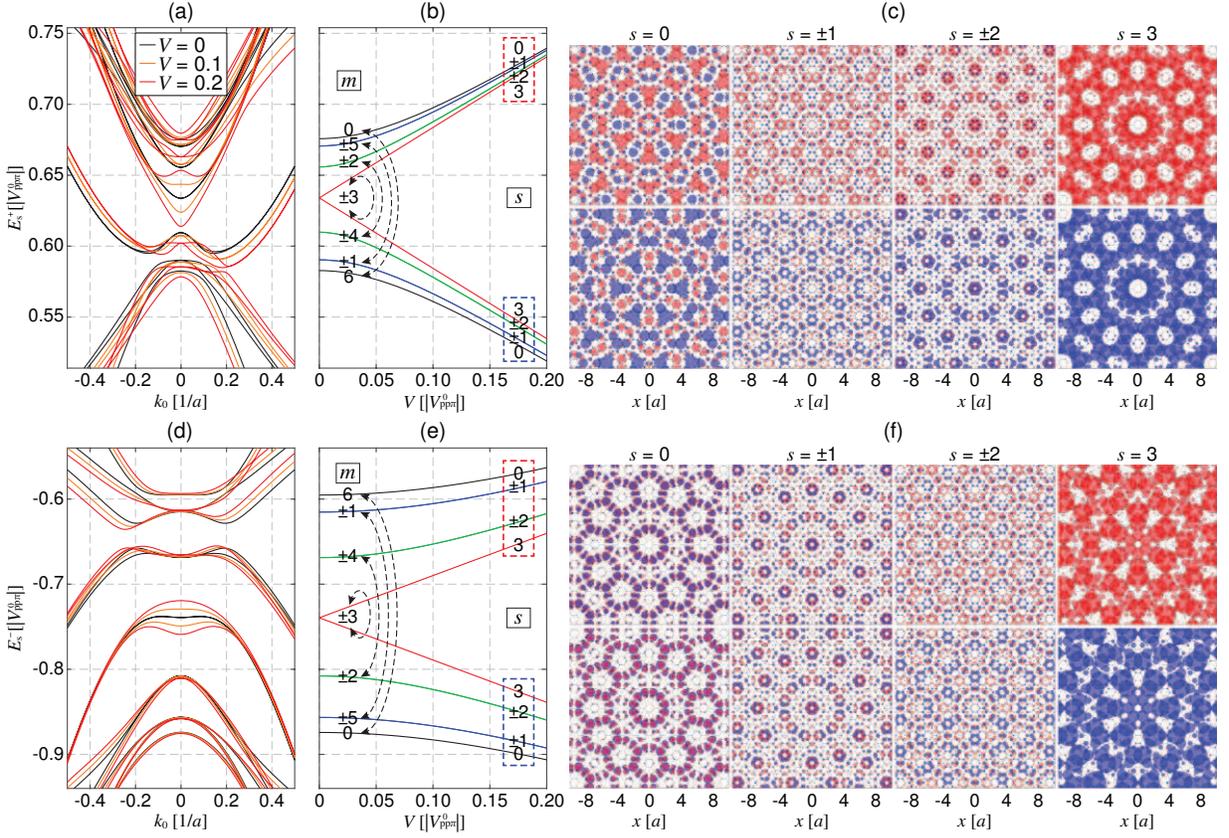}
	\end{center}
	\caption{
		(a) Conduction band dispersion near $\Vec{C}_n$
		of dodecagonal vdW-QCs under three different interlayer potential asymmetry values, $V=0$, $0.1\,|V_{pp\pi}^0|$, $0.2\,|V_{pp\pi}^0|$.
		(b) Conduction band edges at $\Vec{C}_n$ with
		various $V$.
		Indices $m$ and $s$ show
		the angular momentum of the pristine quasicrystalline states
		with 12-fold rotational symmetry
		and that of 6-fold rotational symmetry
		under interlayer potential asymmetry.
		Dashed arrows show the interaction
		between the constituent quasicrystalline states by $\mathcal{H}_V$.
		(c) Plots similar to Fig.~\ref{fig_dodeca_bs}(c)
		for $V\ne0$ in the conduction band.
		The top and bottom panels show the LDOS
		of the upper and lower bands, respectively.
		(d), (e), (f): Plots similar to (a), (b), (c)
		for valence band.
		}
	\label{fig_dodeca_bs_w_U}
\end{figure*}

In addition to the quasicrystalline 12-wave mixing,
Fig.~\ref{fig_dodeca_DOS_effects_of_t0}(b)
shows the features from the 2-wave mixing.
The two DOS peaks and the dips in between, which originate at
$E=-0.5\,|V_{pp\pi}^0|$ when $t_0=0$,
show the band edges and pseudogaps
arising from intralayer 2-wave mixing
\cite{koshino2015incommensurate,yao2018quasicrystalline,moon2019quasicrystalline}.
We plot the band opening associated the interlayer and intralayer mixing
as the blue and red arrows
in Figs.~\ref{fig_dodeca_bs}(a) and \ref{fig_dodeca_DOS_effects_of_t0}(a), respectively,
and visualize these interactions
in Fig.~\ref{fig_dodeca_DOS_effects_of_t0}(c).
The band opening
by the 2-wave mixing in the conduction band
is much smaller than that in the valence band
for the same reason as the band opening
via the 12-wave mixing (Sec.~\ref{sec:dodeca_band_structure}).
Although the interlayer interaction strength $t(\Vec{q})$
involved in the intralayer 2-wave mixing
is about 1.50 times the interaction strength
in the interlayer 2-wave mixing,
the intralayer mixing exhibits
smaller band opening
since it involves two successive interlayer interaction
between the states with different energies.
At sufficiently large $|t_0|$ ($>0.08\,|V_{pp\pi}^0|$),
however,
the intralayer interaction gives band opening
throughout the entire Brillouin zone in the valence band
except in the vicinity of the quasicrystalline states.
This means  that the quasicrystalline states
with $m=6,\pm1$ are easily observable in the specific energy window where the quasicrystalline states remain after the weakly coupled bands become gapped.
At $t_0 \rightarrow 0$ limit,
the interlayer 2-wave mixing
emerges at an energy that is the same as the 12-wave mixing,
while the intralayer 2-wave mixing
emerges at an energy much closer to $E=0$.
Again, the states and band opening arising from
these three different mixings
are continuously connected to each other
in the Brillouin zone
[Figs.~\ref{fig_dodeca_DOS_effects_of_t0}(a) and (c)].
The 2-wave mixing can occur in
bilayer hexagonal lattices stacked at any rotation angle $\theta$,
but $\alpha$ and $\beta$ occur
at different energies when $\theta\ne30^\circ$.

\subsubsection{\label{sec:dodeca:H_V}Effects of interlayer potential asymmetry}


Figures \ref{fig_dodeca_bs_w_U}(a) and (d) show the dispersion
in the conduction band and valence band
near $\Vec{C}_n$ of dodecagonal vdW-QCs
under three different interlayer asymmetric potential,
$V=0$, $0.1$, $0.2$ in units of $|V_{pp\pi}^0|$.
Again, the states with $m=\pm3$ at $\Vec{k}_0=\Vec{0}$
lose their degeneracy
and all of the band edges, in both the conduction band and valence band,
move away from $E_{m=\pm3}$ as $V$ increases.

At $\Vec{k}_0=\Vec{0}$,
the interlayer potential asymmetry couples
the eigenstates of
$\mathcal{H}_{\mathrm{ring}}$,
$\Vec{v}_m^b$ ($b$ is $+$ for conduction band and $-$ for valence band),
whose angular momenta differ by $\pm6$,
\begin{eqnarray}
&&\langle \Vec{v}_{m'}^{b'} | \mathcal{H}_V | \Vec{v}_m^b \rangle \nonumber\\
&&=
\left\{ 
\begin{array}{cl}
 -\frac{V}{2} (c_{m',1}^{b'}c_{m,1}^b+c_{m',2}^{b'}c_{m,2}^b)
& (m-m' \equiv 6\,(\mathrm{mod}\,12)), \\
 0 & (\mathrm{otherwise}),
\end{array}
\right.
\end{eqnarray}
since the diagonal elements of $\mathcal{H}_V$
work as a staggered potential
with 1/6 period of the ring
in the dual tight-binding lattice.
Thus, the Hamiltonian matrix
is reduced to six $4\times 4$ matrices
\begin{eqnarray}
\tilde{\mathcal{H}}_{m,m'}&&=
\begin{pmatrix}
\mathcal{C}_m & -\frac{V}{2}R(\frac{\phi_m-\phi_{m'}}{2}) \\
-\frac{V}{2}R^{-1}(\frac{\phi_m-\phi_{m'}}{2}) & \mathcal{C}_{m'}
\end{pmatrix}, \nonumber\\
\mathcal{C}_m&&=
\begin{pmatrix}
E_m^- & 0 \\ 0 & E_m^+
\end{pmatrix},
\label{eq_dodeca_H_E_of_U}
\end{eqnarray}
in the bases of
$(\Vec{v}_m^-,\Vec{v}_m^+,\Vec{v}_{m'}^-,\Vec{v}_{m'}^+)$
for $(m,m')=(0,6),(1,-5),(2,-4),(3,-3),(4,-2),(5,-1)$,
where $R(\phi)$ is a rotation matrix.
The electronic states lose the 12-fold rotational symmetry,
since $\mathcal{H}_V$ lifts the degeneracy of $|\Vec{k}^{(n)}\rangle$
in different layers,
and are characterized by 
the angular quantum number 
$s\,(=0,\pm1,\pm2,3)\equiv m \equiv m'\,(\mathrm{mod}\,6)$
for a 6-fold rotational symmetry.
We obtain two band edges $E_s^+$
in the conduction band and another two $E_s^-$
in the valence band
by diagonalizing Eq.\,\eqref{eq_dodeca_H_E_of_U}.
These are plotted against $V$ in Figs.~\ref{fig_dodeca_bs_w_U}(b) and (e),
for the conduction and valence bands, respectively.
It is straightforward to show that
the states with a quantum number $s$ and $-s$ ($s=1,2$) are degenerate, since
the reduced Hamiltonian satisfies
\begin{eqnarray}
&& \tilde{\Sigma}^{-1} \tilde{\mathcal{H}}_{m,m'} \tilde{\Sigma}
= \tilde{\mathcal{H}}_{-m,-m'}, \nonumber\\
&& \tilde{\Sigma} = \mathrm{diag}(1,-1,1,-1).
\end{eqnarray}
In most practical cases,
Eq.\,\eqref{eq_dodeca_H_E_of_U} can be further reduced to
two $2\times 2$ matrices,
since the interaction between the state in
the conduction band and 
the state in valence band is almost negligible. This is partly
due to the large energy difference
and partly due to $(\phi_m-\phi_{m'})/2 \approx 0$
(since $\phi_m \approx \phi_{m'} \approx 90^\circ$).

The dashed arrows in Figs.~\ref{fig_dodeca_bs_w_U}(b) and (e)
show the interaction
between quasicrystalline states between $\Vec{v}_m^\pm$ and $\Vec{v}_{m'}^\pm$
by $\mathcal{H}_V$.
The states in the conduction band exhibit
larger mixing
between the constituent quasicrystalline states
than those in the valence band,
due to the smaller energy difference (not shown).
And, similar to the octagonal vdW-QCs [Fig.~\ref{fig_octa_bs_w_U}(c)],
materials with weaker $|t_0|$ under larger $V$
experience larger energy shift, mixing,
and accordingly larger spatial layer-polarization
because of Eq.~\eqref{eq_layer_polarization}
(Appendix \ref{sec:App:dodeca_energy_span}).
Again, the states with $s=3$ are special in that
the constituent states $\Vec{v}_{m=3}^\pm$
and $\Vec{v}_{m=-3}^\pm$
are always fully mixed
regardless of the values of $t_0$ and $V$,
due to the degeneracy between
$E_{m=3}^\pm$ and $E_{m=-3}^\pm$.

We plot the LDOS of the states in the conduction band and valence band
with $s=0,\pm1,\pm2,3$ at $V=0.2\,|V_{pp\pi}^0|$
in Figs.~\ref{fig_dodeca_bs_w_U}(c) and (f), respectively,
where the top and bottom panels in each figure show the LDOS of
the upper and lower bands, respectively.
Again, the stronger the mixing,
the more the wave functions are layer polarized,
and the wave functions $\Psi_{s=3}^\pm$ are mostly
polarized to either layer
even at very weak $V$.
The LDOS profile of each layer-polarized state
is exactly consistent with the profile of each layer
in the absence of the potential asymmetry
[Fig.~\ref{fig_dodeca_bs}(c)],
since $\mathcal{H}_V$ does not change the Umklapp scattering paths.


In dodecagonal vdW-QCs with sublattice symmetry (i.e., $\Delta=0$), an
interlayer potential asymmetry
does not open a gap at the Dirac point.
This is because the coexistence of the time reversal symmetry
and the in-place $C_2$ rotation symmetry
requires vanishing of the Berry curvature
at any nondegenerate point the the energy band \cite{moon2014optical},
and this guarantees the robustness of band touching points
in two-dimensional systems \cite{PhysRevB.88.115409}.
Just like twisted bilayer graphene with any rotation angle
\cite{moon2012energy},
dodecagonal vdW-QCs composed of two hexagonal lattices
with $\Delta=0$
has the $C_2$ symmetry,
even in the presence of
interlayer potential asymmetry because $C_2$ does not flip the layers.
Thus, the Dirac points of dodecagonal vdW-QCs with $\Delta=0$
are protected even in the presence of interlayer potential asymmetry.

\subsubsection{Effects of sublattice potential asymmetry}

\begin{figure*}
	\begin{center}
		\leavevmode\includegraphics[width=0.7\hsize]{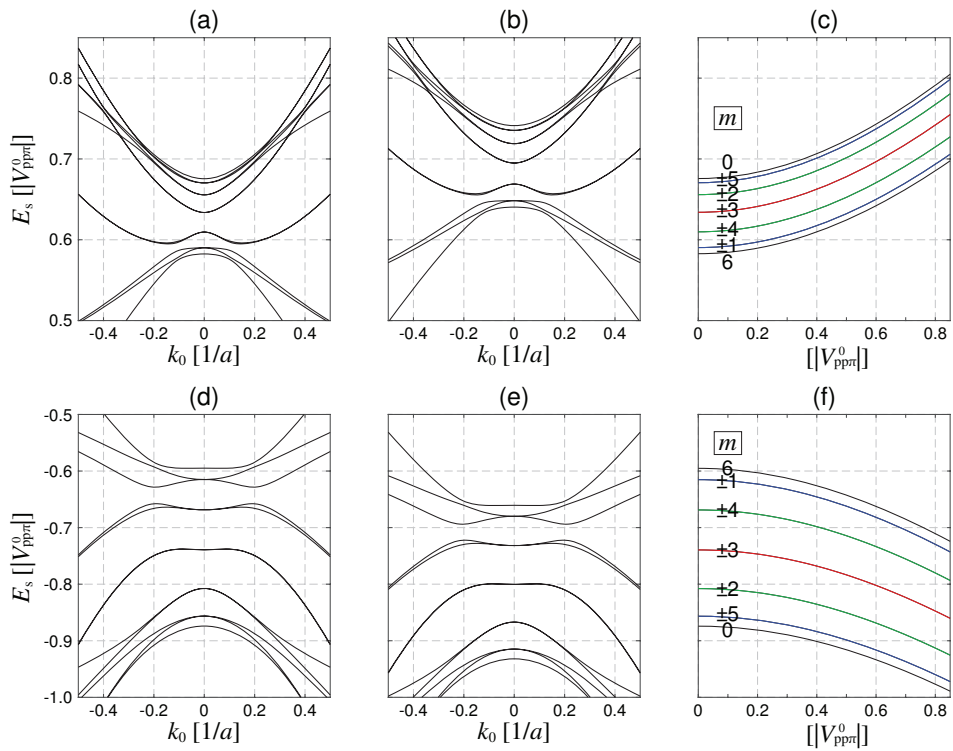}
	\end{center}
	\caption{
		(a) and (b): Conduction band dispersion
		near $\Vec{C}_n$ of a dodecagonal vdW-QC
		with sublattice potential asymmetry of 
		(a) $\Delta = 0$ and (b) $-0.60\,V_{pp\pi}^0$, respectively.
		(c) Band edges at $\Vec{C}_n$ with
		various $\Delta$.
		Indices $m$ show the angular momentum
		of the pristine quasicrystalline states
		with 12-fold rotational symmetry.
		(d), (e), (f): Plots similar to (a), (b), (c)
		for valence band.
		}
	\label{fig_dodeca_bs_w_Delta}
\end{figure*}

We plot the conduction band and valence band
near $\Vec{C}_n$ of a dodecagonal vdW-QC
with sublattice potential asymmetry of
$\Delta = -0.60\, V_{pp\pi}^0$
in Figs.~\ref{fig_dodeca_bs_w_Delta}(b) and (e), respectively,
and plot the bands in the absence the asymmetry
in (a) and (d) as a reference.
$\Delta\ne0$ in the current model
makes a band opening as large as $\Delta$
at the energy range
centered at $E=0$.
Unlike the interlayer potential asymmetry, however,
we can see that
breaking the sublattice symmetry does not make
dramatic change to
the band structures near the quasicrystalline states.

The potential which breaks the sublattice symmetry, $\mathcal{H}_\Delta$,
couples the eigenstates of
$\mathcal{H}_{\mathrm{ring}}$,
$\Vec{v}_m^b$,
whose angular momenta differ by $\pm3$,
\begin{eqnarray}
&&\langle \Vec{v}_{m'}^{b'} | \mathcal{H}_\Delta | \Vec{v}_m^b \rangle \nonumber\\
&&=
\left\{ 
\begin{array}{cl}
  \frac{(1-i)\Delta}{4} (c_{m',1}^{b'}c_{m,1}^b-c_{m,2}^{b'}c_{m,2}^b)
  & (m-m' \equiv -3\,(\mathrm{mod}\,12)), \\
  \frac{(1+i)\Delta}{4} (c_{m',1}^{b'}c_{m,1}^b-c_{m,2}^{b'}c_{m,2}^b)
  & (m-m' \equiv 3\,(\mathrm{mod}\,12)), \\
0 & (\mathrm{otherwise}),
\end{array}
\right.
\end{eqnarray}
since the diagonal elements of $\mathcal{H}_\Delta$
work as a potential
with 1/3 period of the ring
in the dual tight-binding lattice.
Thus, $\mathcal{H}_\Delta$ couples
quasicrystalline states with four different $m$,
$(m_1,m_1',m_2,m_2')=(0,3,6,-3),(1,4,-5,-2),(-1,2,5,-4)$,
and the Hamiltonian matrix
is reduced to three $8\times 8$ matrices
\begin{eqnarray}
\tilde{\mathcal{H}}_{m_1,m_1',m_2,m_2'}&&=
\left(
\begin{array}{cccc}
    \mathcal{C}_{m_1} & \mathcal{D}_{m_1,m_1'} & 0 & \mathcal{D}_{m_2',m_1}^\dagger \\
    \mathcal{D}_{m_1,m_1'}^\dagger & \mathcal{C}_{m_1'} & \mathcal{D}_{m_1',m_2} & 0 \\
    0 & \mathcal{D}_{m_1',m_2}^\dagger & \mathcal{C}_{m_2} &
    \mathcal{D}_{m_2,m_2'} \\
    \mathcal{D}_{m_2',m_1} & 0 & \mathcal{D}_{m_2,m_2'}^\dagger &
    \mathcal{C}_{m_2'}
\end{array}
\right), \nonumber\\
\mathcal{D}_{m_a,m_b}&&=\frac{(1+i)\Delta}{4}
\begin{pmatrix}
\cos \bar{\phi}_m &
\sin \bar{\phi}_m \\
\sin \bar{\phi}_m &
-\cos \bar{\phi}_m
\end{pmatrix}
\label{eq_dodeca_H_E_of_Delta}
\end{eqnarray}
in the bases of
$(\Vec{v}_{m_1}^-,\Vec{v}_{m_1}^+,\Vec{v}_{m_2}^-,\Vec{v}_{m_2}^+,\Vec{v}_{m_1'}^-,\Vec{v}_{m_1'}^+,\Vec{v}_{m_1'}^-,\Vec{v}_{m_1'}^+)$,
where $\bar{\phi}_m = \equiv (\phi_{m_a}+\phi_{m_b})/2$.
The electronic states lose the 12-fold rotational symmetry,
and are characterized by 
the angular quantum number 
$s\,(=0,\pm1)\equiv m_i ,(\mathrm{mod}\,3)$
($m_i=m_1,m_2,m_1',m_2'$)
for a 3-fold rotational symmetry.
Again,
the states with a quantum number $s=1$ and $s=-1$ are degenerate,
due to the symmetry of the Hamiltonian.
And in most practical cases,
Eq.\,\eqref{eq_dodeca_H_E_of_Delta} can be further reduced to
four $2\times 2$ matrices,
since the interaction between the state in
the conduction band and 
the state in valence band is almost negligible
due to the large energy difference.

Figures \ref{fig_dodeca_bs_w_Delta}(c) and (f) show
the energies $E_s^\pm$ in the unit of $|V_{pp\pi}^0|$
plotted against $\Delta$.
Unlike the interlayer potential asymmetry,
$\Delta$ merely shifts the energies slightly away from the Dirac point
and does not make dramatic change to the quasicrystalline states,
which is consistent with the band structures
in Figs.~\ref{fig_dodeca_bs_w_Delta}(b) and (e).
This is because,
$|c_{m,1}^b|\approx|c_{m,2}^b|$
in most practical systems with $|t_0| < h_0/5$
($\phi_m \approx 90^\circ$)
at this high energy regime.
Thus, the potential, which has the opposite sign
between the sublattices,
is almost cancelled in the intraband interaction ($b = b'$),
due to the phase cancellation.
On the other hand, although
materials with much higher $|t_0|$
have finite contribution from the sublattice phases
in the intraband interaction,
the overall interaction
is still very weak
since the energy difference
between the quasicrystalline states
increases as $|t_0|$ grows.
In any system,
the interband interaction ($b \ne b'$)
is always negligible
due to the large energy difference.
Thus,
a bilayer of hexagonal lattices with sublattice potential asymmetry
stacked at $30^\circ$
will also exhibit the quasicrystalline states
and DOS
analogous to the quasicrystals
composed of bilayer graphene.
If the two layers have different $\Delta$,
other than a simple sign difference,
the degeneracy of the $m=\pm3$ states is lifted
but the other states remain almost the same.


\section{\label{sec:conclusions}Conclusions}

We investigated the electronic structures of
quasicrystals composed of the incommensurate stack of
atomic layers (vdW-QCs)
for every rotational symmetry possible in two-dimensional space.
We show that the rotational symmetry of the quasicrystal
as well as the translational symmetries
of the constituent atomic layers
give the quasicrystalline resonant coupling
between the intrinsic states of the constituent layers.
Furthermore, we reveal the emergence of the quasi-band dispersion and wave functions
respecting the quasicrystalline order of each system.
Although the quasicrystalline states
coexist in energy with weakly coupled states
(e.g., the states arising from the interaction
which is typically known as \textit{moir\'{e} interaction})
in general,
we showed that some quasicrystalline states, which are usually obscured by additional weakly coupled states, are more prominent in quasicrystals with strong interlayer interaction.

From the analysis on the symmetry of the interlayer interaction,
we show that even the atomic layers
with different types of orbitals
will also exhibit the quasicrystalline states
if all the dominant interlayer interaction
occurs between the atomic orbitals
having the same magnetic quantum number.
In this sense,
most of dodecagonal vdW-QCs including those composed of
transition metal dichalcogenides
will also clearly show the quasicrystalline states.

Besides, we investigate 
the effects of lifting both interlayer and sublattice symmetry
on the electronic structures.
Since the quasicrystalline order arises
from the resonant interaction between the states in both layers,
we can switch between the states fully respecting quasicrystal symmetry
and those satisfying only half the symmetry
by turning on and off the interlayer symmetry.
We also analytically interpret the mixing between the quasicrystalline states,
which may influence other physical properties such as optical selection rules.
The quasicrystalline states in the middle are special in that
they are always fully mixed
and $100\%$ layer polarized
regardless of the magnitude of the interlayer asymmetry.
On the other hand, we show that sublattice potential asymmetry
in hexagonal lattices
does not make a dramatic difference - it results only in a constant energy shift from the original quasicrystalline states.

This is the first theoretical work
which investigated the formation of quasicrystalline states
for every possible two-dimensional bilayer quasicrystal system,
which will lead to extended
exploration of rich quasicrystal physics in
designer quasicrystals.
In experiment, the quasicrystalline bands
will be observed in Angle-resolved photoemission spectroscopy
clearly at $C_n$
then much weakly at other wave vectors including $\Gamma$
by Umklapp scattering.
Optical selection rule would be different from that of
the constituent layers,
and we can switch between the two different rules
by switching on and off the interlayer potential asymmetry.
And, unlike the moir\'{e} superlattices stacked
at other angles,
vdW-QCs will not show circular dichroism.
Besides, scanning tunneling microscope will show
the LDOS respecting the quasicrystalline tiling.
In the most widely studied system, graphene quasicrystals,
however, most quasicrystalline states
coexist in energy with almost decoupled monolayer states.
Thus, the quasicrystalline pattern of LDOS
will be visible as a weak deviation from the uniform electron distribution
or by exposing the states
by increasing the interlayer interaction,
e.g., by applying pressure, by intercalation of ions,
or by addition of barrier atomic layers.
Besides, the growth of the band flatness near $C_n$
with respect to the increase of the interlayer interaction
will significantly
reduces the conductivity, and the electron-electron interaction
in such flat bands may serve as the source of many interesting
phenomena such as the enhancement of electron-phonon coupling.

\textit{Note added.}
During the completion of this work,
we became aware of recent
theoretical works on the pressure and electric field dependence
of quasicrystalline electronic states in $30^\circ$ twisted bilayer graphene
\cite{yu2020pressure}.

\begin{acknowledgments}
This work was supported by
Science and Technology Commission of Shanghai Municipality
grant no.~19ZR1436400, and
NYU-ECNU Institute of Physics at NYU Shanghai.
This research was carried out on the High Performance Computing resources
at NYU Shanghai.

\end{acknowledgments}

\appendix


\section{Quasicrystalline states with weaker interaction}
\label{sec:App:weaker_resonant_condition}
\begin{figure*}
	\begin{center}
		\leavevmode\includegraphics[width=0.9\hsize]{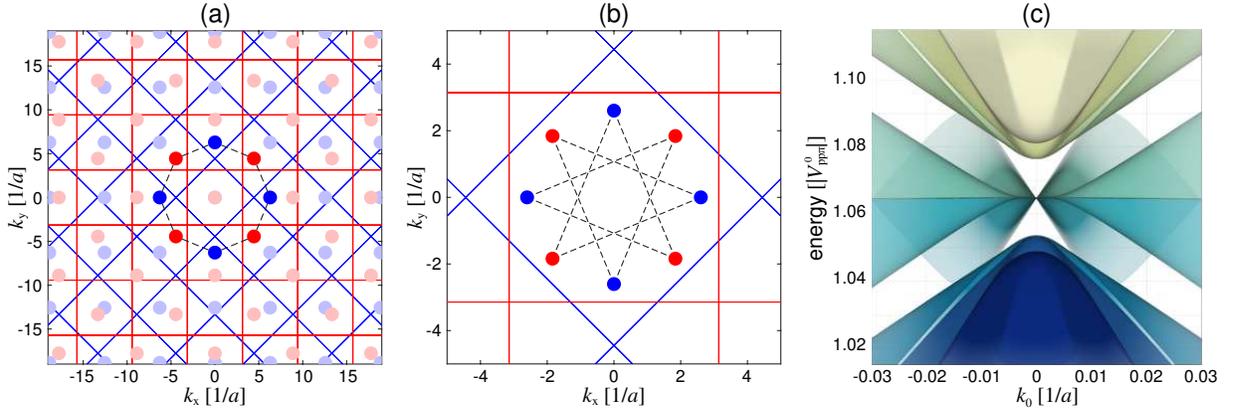}
	\end{center}
	\caption{
		(a) and (b) Plots similar to Figs.~\ref{fig_atomic_structure_and_BZ}(b) and (c)
		for the next strongest quasicrystalline interaction
		in octagonal vdW-QCs,
		where $\hat{\Vec{k}}=\Vec{0}$.
		(c) Electronic structures near the second dominant quasicrystalline states
		of octagonal vdW-QCs
		calculated by the 8-ring effective model.
		}
	\label{Fig_octa_bs_next_strongest}
\end{figure*}

As mentioned in Sec.~\ref{sec:RND:resonant_condition},
the sets of the waves in Figs.~\ref{fig_atomic_structure_and_BZ}(b) and (e)
are not the only set of states
which show a resonant coupling in each system.
We can find more sets of states, with different wave numbers,
showing the resonant interaction
respecting the rotational symmetry of the quasicrystals.
For example, the eight states $\Vec{k}$ (red) and $\tilde{\Vec{k}}'$ (blue)
in Fig.~\ref{Fig_octa_bs_next_strongest}(a)
also form a circular chain in the dual tight-binding lattice.
Note that $\hat{\Vec{k}}$ for these states ($=\Vec{0}$)
is different from that for the states in Fig.~\ref{fig_atomic_structure_and_BZ}(b).
These states are mapped to
$\Vec{k}$ (red) and $\tilde{\Vec{k}}$ (blue)
in the first Brillouin zone,
experience a resonant interaction,
and form quasicrystalline states.
Figure \ref{Fig_octa_bs_next_strongest}(c) shows
the band structures near the quasicrystalline states
arising from these eight states.
It should be noted that, however,
the strength of the resonant interaction, $|t(\Vec{q})|$,
for the states in Fig.~\ref{Fig_octa_bs_next_strongest}(a)
is much weaker than that for the states in Fig.~\ref{fig_atomic_structure_and_BZ}(b).
This is because $|t(\Vec{q})|$ decays fast as $|\Vec{q}|$ grows,
and the former states have the chain with a longer segment length (=$|\Vec{q}|$).
Thus, the band opening in Fig.~\ref{Fig_octa_bs_next_strongest}(c)
is much smaller than that in Figs.~\ref{fig_octa_bs}(a) and (b).
Dodecagonal vdW-QCs also have more sets of states
showing the resonant interaction.
In most systems, however,
such states can be mostly neglected
since their interaction strengths are very weak,
and they are also mixed with other types of interaction (e.g., 2-wave mixing).
Thus, the sets in Figs.~\ref{fig_atomic_structure_and_BZ}(b)
and (e) give the strongest interaction,
i.e., largest energy separation and clear quasicrystalline order,
since these states form the rings with the shortest
distance between neighboring states
in the dual tight-binding lattices.

\section{Particle-hole symmetry of the Hamiltonian of octagonal quasicrystal}
\label{sec:App:octa_k1st}

By considering only the nearest neighbor pairs in the intralayer interaction,
the ring Hamiltonian $\mathcal{H}_{\mathrm{ring}}$
of the octagonal vdW-QCs [Eq.\,\eqref{eq_octa_H_ring}],
up to the first order to $\Vec{k}_0 = (k_{0,x},k_{0,y})$,
can be transformed to
\begin{widetext}
\begin{equation}
U^{-1} \mathcal{H}_{\mathrm{ring}}(\Vec{k}_0) U = \mathcal{H}_{\mathrm{ring}}'(\Vec{k}_0)=
\begin{pmatrix}
H'^{(-3)} & C^* &&&& C \\
C & H'^{(-2)} & C^* \\
& C & H'^{(-1)} & C^* \\
&& \ddots & \ddots & \ddots \\
&&& C & H'^{(3)} & C^* \\
C^* &&&& C & H'^{(4)}
\end{pmatrix}
\end{equation}
\end{widetext}
with a transformation matrix
\begin{equation}
U=
(\Vec{v}_{-3},\Vec{v}_{-2},\Vec{v}_{-1},\cdots,\Vec{v}_{3},\Vec{v}_{4}),
\end{equation}
where
$\Vec{v}_m=\frac{1}{\sqrt{8}}(\mu_m^{-3},\mu_m^{-2},\cdots,\mu_m^{3},\mu_m^4)^\mathrm{T}$
($\mu_m=e^{i 5 \pi m /4}$)
is the eigenstate of the quasicrystalline state
with a quantized angular momentum of $m$, 
$H'^{(m)}=h_0-2t_0 \cos(5 \pi m /4)$, and
$C(\Vec{k}_0)= \sin(\sqrt{2}\pi) a V_{pp\pi}^0 (k_{0,x}-ik_{0,y})$.
Then, it is straightforward to show that $\mathcal{H}'_{\mathrm{ring}}$
has a particle-hole symmetry with respect to the energy $E=h_0$,
\begin{eqnarray}
&&\Sigma^{-1} (\mathcal{H}_{\mathrm{ring}}'-h_0 \mathbb{I}) \Sigma
= -(\mathcal{H}_{\mathrm{ring}}'-h_0 \mathbb{I}), \nonumber\\
&&\Sigma=
\begin{pmatrix}
&& \sigma_z \\
&&& \sigma_z \\
\sigma_z \\
& \sigma_z
\end{pmatrix},
\end{eqnarray}
where $\mathbb{I}$ is an $8\times8$ unit matrix.
This immediately demonstrates that
if $\Psi(\Vec{k}_0)$ is an eigenstate of $\mathcal{H}_{\mathrm{ring}}'$
with an energy of $h_0 + E$,
then $\Sigma^{-1} \Psi(\Vec{k}_0)$
is an eigenstate of energy $h_0 - E$.


\section{\label{sec:App:dodeca_band_edges}Band edges of quasicrystalline states of dodecagonal vdW-QCs with various $t_0$}

\begin{figure}
	\begin{center}
		\leavevmode\includegraphics[width=0.7\hsize]{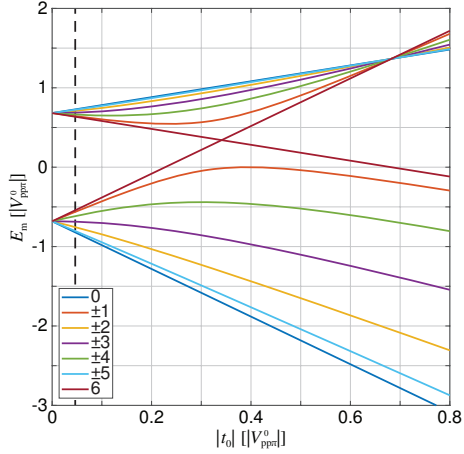}
	\end{center}
	\caption{
		Band edges of quasicrystalline states of dodecagonal
		vdW-QCs [Eq.~\eqref{eq_dodeca_Em}]
		with various interlayer interaction strength $t_0$.
		The black dashed line corresponds to the band edges
		for the system considered in Fig.~\ref{fig_dodeca_bs},
		and the numbers show the quantized angular momentum $m$.
		}
	\label{fig_dodeca_t0}
\end{figure}

Figure \ref{fig_dodeca_t0} shows the band edges
of quasicrystalline states of dodecagonal vdW-QCs
[Eq.~\eqref{eq_dodeca_Em}] with various interlayer
interaction strength $t_0$
up to the strong interaction regime.
We can see that the energy spacing
between the quasicrystalline states
increases as $|t_0|$ increases in
most practical interaction strength,
i.e., $|t_0| < h_0/2$ ($h_0=-0.682\,V_{pp\pi}^0$),
while the energy spacing in the conduction band decreases as $|t_0|$ exceeds
$h_0/2$.

\section{\label{sec:App:dodeca_energy_span}Energies and wave functions of dodecagonal quasicrystal with interlayer potential asymmetry}

Figures \ref{fig_dodeca_bs_w_U}(b) and (e)
show that the states with $s=0$
determine the energy span of the resonant states
in the presence of the interlayer potential asymmetry,
in both the conduction band and valence band.
Equation \eqref{eq_dodeca_H_E_of_U} shows the coupling
between the quasicrystalline states
with angular momentum $m$
by interlayer potential asymmetry.
In most practical cases,
the matrix can be further reduced to
two $2\times 2$ matrices,
\begin{equation}
\tilde{\mathcal{H}}^\pm_{m,m'}=
\begin{pmatrix}
E_m^\pm & -V/2 \\
-V/2 & E_{m'}^\pm
\end{pmatrix},
\end{equation}
for the conduction band ($\tilde{\mathcal{H}}^+_{m,m'}$)
and valence band ($\tilde{\mathcal{H}}^-_{m,m'}$),
since the interaction between the state in
the conduction band and 
the state in valence band is almost negligible
due to the large energy difference.
For the coupled states with $s=0$ (i.e., $m=0$ and $m'=6$),
the interaction between the conduction band and
valence band is completely forbidden
due to the sublattice symmetry.
Then, we get $E_{s=0}^-=-h_0\pm \sqrt{9t_0^2+V^2/4}$ in the valence band
and $E_{s=0}^+=h_0\pm \sqrt{t_0^2+V^2/4}$ in the conduction band,
in the most practical systems with $|h_0| > 2|t_0|$.
Thus, the states in the conduction band exhibit
smaller energy span than those in the valence band.

The wave functions of the higher energy states
in both the conduction band and valence bands
are 
$\Psi_{s=0}^\pm = \sin (\tilde{\phi}/2) \Vec{v}_{m=0}^\pm + \cos (\tilde{\phi}/2) \Vec{v}_{m=6}^\pm$,
and the lower energy states are
$\Psi_{s=0}^\pm = \cos (\tilde{\phi}/2) \Vec{v}_{m=0}^\pm - \sin (\tilde{\phi}/2) \Vec{v}_{m=6}^\pm$,
where $\tilde{\phi}$ is $\tan^{-1}(-V/(6t_0))$ for valence band
and $\tan^{-1}(V/(2t_0))$ for conduction band.
As $\tilde{\phi}$ becomes close to $90^\circ$,
i.e., in materials with smaller $|t_0|$ and $|V|>>|t_0|$,
$\Psi_{s=0}^\pm$ becomes $(1/\sqrt{2})(\Vec{v}_0^\pm+\Vec{v}_6^\pm)$
for the upper state and
$(1/\sqrt{2})(\Vec{v}_0^\pm-\Vec{v}_6^\pm)$
for the lower state.
Since
$\Vec{v}_0^\pm=(1/\sqrt{24})(1,1,1,1,\cdots,1)\bigotimes(1,\pm1)$ and
$\Vec{v}_6^\pm=(1/\sqrt{24})(-1,1,-1,1,\cdots,1)\bigotimes(1,\pm1)$,
$\Psi_{s=0}^\pm$ becomes polarized to either layer,
i.e., the state is mostly composed 
of the Bloch bases
with $n$ of even (layer 1) or odd (layer 2) numbers,
in the systems with small interlayer interaction strength $|t_0|$.
On the other hand,
the states with $s=3$ are always
$100\%$ polarized to either layer.

\bibliography{qc}

\end{document}